\documentclass{article}

\usepackage[T1]{fontenc}
\usepackage[latin9]{inputenc}
\usepackage{float}
\usepackage{rotfloat}
\usepackage{amsmath}
\usepackage{amssymb}
\usepackage{graphicx}
\usepackage{caption}
\usepackage{subcaption}
\usepackage{booktabs}
\usepackage{mathtools}
\usepackage{esint}
\usepackage{bbm}
\usepackage{color}
\usepackage[round,authoryear]{natbib}
\usepackage{bm}
\usepackage[shortlabels]{enumitem}
\usepackage{amsthm}  
\usepackage{float}
\usepackage{nicefrac} 
\usepackage{comment}
\usepackage{tikz}
\usepackage[makeroom]{cancel}
\usepackage{ulem}

\usepackage{xcolor,colortbl}
\def\checkmark{\tikz\fill[scale=0.4](0,.35) -- (.25,0) -- (1,.7) -- (.25,.15) -- cycle;} 

\usetikzlibrary{shapes.misc}

\tikzset{cross/.style={cross out, draw=black, minimum size=2*(#1-\pgflinewidth), inner sep=0pt, outer sep=0pt},
cross/.default={1pt}}

\newcommand{\Cross}{$\mathbin{\tikz [x=1.4ex,y=1.4ex,line width=.2ex, red] \draw (0,0) -- (1,1) (0,1) -- (1,0);}$}%
\newcommand{\Checkmark}{$\color{black!30!green}\checkmark$}

\newcommand{\possible}{\cellcolor{green!10}\Checkmark}
\newcommand{\notpossible}{\cellcolor{red!10}\Cross}

\usetikzlibrary{positioning}
\usetikzlibrary{arrows}
\usepackage{lscape}
\usepackage{tcolorbox}
\definecolor{myviolet}{rgb}{0.73,0.56,0.64}
\newtcolorbox{mybox}{
    arc=0pt,
    boxrule=0pt,
    colback=myviolet,
    width=.2\textwidth,   
    colupper=white,
    fontupper=\bfseries
}

\defcitealias{gagliardini2016time}{GOS}

\tikzset{block/.style={minimum width=1.1 em, minimum height= 1em, rounded corners= 4pt}}

\DeclareMathOperator*{\subG}{subG}
\DeclareMathOperator*{\plim}{plim}
\DeclareMathOperator*{\argmin}{argmin}
\DeclareMathOperator*{\tr}{tr}
\DeclareMathOperator*{\vech}{vech}
\DeclareMathOperator*{\vect}{vec}
\DeclareMathOperator*{\supp}{supp}
\DeclareMathOperator*{\cov}{Cov}
\DeclareMathOperator*{\diag}{diag}

\def\betaone{\beta_{1,i}}
\def\betatwo{\beta_{2,i}}

\DeclareRobustCommand{\model}{\ensuremath{\mathcal{M}}}
\DeclareRobustCommand{\groupset}{\ensuremath{\mathcal{G}}}
\DeclareRobustCommand{\group}{\ensuremath{g}}
\DeclareRobustCommand{\real}{\ensuremath{\mathbb{R}}}
\DeclareRobustCommand{\natural}{\ensuremath{\mathbb{N}}}
\DeclareMathOperator{\eig}{eig}

\usepackage[hyperindex=true,pdftitle={},
pdfauthor={Gaetan Bakalli},colorlinks=TRUE,
pagebackref=false,citecolor=blue,plainpages=false,
pdfpagelabels]{hyperref}

\def\boxit#1{\vbox{\hrule\hbox{\vrule\kern3pt
          \vbox{\kern3pt#1\kern3pt}\kern3pt\vrule}\hrule}}

\usepackage{framed}
\makeatletter
 {\par\unskip\endMakeFramed%
 \at@end@of@kframe}
\makeatother

\newtheoremstyle{mytheoremstyle} 
    {0.3cm}                      
    {0.3cm}                        
    {\itshape}                   
    {}                           
    {\scshape}                   
    {: }                          
    {0em}                       
    {}  

\theoremstyle{mytheoremstyle}

\newtheorem{Lemma}{Lemma}

\newtheorem{Proposition}{Proposition}

\newtheorem{Aassumption}{Assumption}

\newtheorem{Bassumption}{Assumption}

\newtheorem{RRestriction}{Restriction}

\newtheoremstyle{myExampleRemarkstyle} 
    {0.3cm}                    
    {0cm}                           
    {\itshape}                   
    {}                           
    {\scshape}                   
    {: }                          
    {0em}                       
    {}  

\theoremstyle{myExampleRemarkstyle}
 
\newtheoremstyle{simuStyle}
{0.3cm} 
{0cm} 
{} 
{} 
{\bfseries} 
{.} 
{0em} 
{} 

\theoremstyle{simuStyle}

\newtheoremstyle{stratStyle}
{0.3cm} 
{0cm} 
{} 
{} 
{\scshape} 
{: } 
{0em} 
{} 

\usepackage{soul}
\theoremstyle{stratStyle}

\usepackage{times}

\title{A penalized two-pass regression to predict \\ stock returns with time-varying risk premia}

\author{Gaetan Bakalli$^{a}$ and St\'ephane Guerrier$^{b,c}$ and Olivier Scaillet$^{b,d}$}
\date{July 2022}

\begin{document}

\maketitle
\begin{abstract}
\noindent We develop a penalized two-pass regression with time-varying factor loadings. The penalization in the first pass enforces sparsity for the time-variation drivers while also maintaining compatibility with the no-arbitrage restrictions by regularizing appropriate groups of coefficients. The second pass delivers risk premia estimates to predict equity excess returns. Our Monte Carlo results and our empirical results on a large cross-sectional data set of US individual stocks show that penalization without grouping can yield to nearly all estimated time-varying models violating the no-arbitrage restrictions. Moreover, our results demonstrate that the proposed method reduces the prediction errors compared to a penalized approach without appropriate grouping or a time-invariant factor model.  

\end{abstract}

\medskip{}



\medskip{}

\noindent {\scriptsize{}{}$^{a}$Emlyon Business School, $^{b}$Geneva School of Economics and Management, University of Geneva, $^{c}$Faculty of Science, University of Geneva, $^{d}$Swiss Finance
Institute.}{\scriptsize \par}

\section{Introduction}
\label{sec:In}

Under the arbitrage pricing theory \citep{ross2013arbitrage, Chamberlain_Rothschild_1983}, we know that risk premia are drivers of expected excess returns. Hence, estimating them should be useful for prediction of future equity excess returns. The workhorse to estimate equity risk premia in a linear multi-factor setting
is the two-pass cross-sectional regression method developed by \cite{jensen1972capital} and \cite{fama1973risk}. A series of papers address
its large and finite sample properties for linear factor models with time-invariant
coefficients; see, for example, \cite{shanken1985multivariate,shanken1992estimation}, \cite{jagannathan1998asymptotic}, \cite{shanken2007estimating}, \cite{kan2013pricing}, and the review paper of \cite{jagannathan2010analysis} (see \cite{bryzgalova2019bayesian} for a recent Bayesian approach). In a time-varying setting, 
\cite{gagliardini2016time} (henceforth referred as \citetalias{gagliardini2016time})
study how we can infer the dynamics of equity
risk premia from large stock return data sets under conditional linear
factor models (see also \cite{gagliardini2019estimation} for a review of estimation of large dimensional conditional factor models in finance). They show how to explicitly account for the no-arbitrage restrictions relating the time-varying intercept and the time-varying factor loadings when writing the underlying linear regression to be estimated.
In conditional factor models, we quickly loose parsimony in terms of covariates because of the cross-products induced by the no-arbitrage restrictions. \cite{chaieb2020factors} show that a direct application of the \citetalias{gagliardini2016time}
methodology in an international setting is challenging because of the large number of parameters
needed to model the time-variations in factor exposures and risk premia. Applying the
\citetalias{gagliardini2016time} methodology off-the-shelf to an international setting results in few or even zero stocks kept
for several countries. To address this issue, they suggest to rely on iteratively selecting for each stock the most important covariates driving the dynamics of the factor loadings  without violating the no-arbitrage restrictions. 

The aim of this paper is to tackle this issue via LASSO-type penalisation techniques 
\citep{tibshirani1996regression} to enforce sparsity for the time-variation drivers while also maintaining  compatibility with the no-arbitrage restrictions. The shrinkage targets  the time-invariant counterpart of the time-varying models.  In a conditional factor setting, we aim at addressing the ``multidimensional challenge'' of \cite{cochrane2011presidential}, namely select characteristics which really provide independent information about
average excess returns. More specifically, the penalized first-pass (time-series) regression selects and estimates the regression coefficients ensuring a model specification compatible with the no-arbitrage restrictions through the Overlap Group-LASSO (OGL) of \cite{jacob2009group} and its adaptive version of  \cite{percival2012theoretical}, the aOGL, which extends the original Group-LASSO of \cite{yuan2006model} to groups of variables that may overlap.
 Indeed, if we do not introduce a quadratic term (or cross-products) in the time-varying intercept while the covariate is present in the time-varying factor loadings, we introduce \textit{ex-ante} a model with arbitrage  \citep[see (\ref{eq_dgp}) below, and the discussion in ][]{gagliardini2019estimation}. By definition, we cannot estimate a coefficient for which its covariate is absent. On the contrary, if we delete a covariate in the time-varying factor loadings and keep it in the time-varying intercept, then its corresponding coefficients could be shrunk to zero by a standard LASSO for the first-pass regression, and thus could avoid \textit{ex-post} a model with arbitrage if the true model is sparse. In a standard 
Ordinary Least Squares (OLS) first-pass procedure, those time-varying intercept coefficients could be estimated close to zero if the true model does not include that covariate in the time-varying factor loadings. By introducing groups based on finance theory derived from assuming no asymptotic arbitrage opportunities in the economy,  our aOGL approach can only consider models compatible \textit{ex-ante} with the no-arbitrage restrictions by construction. The groups take explicitly into account the links between the time-varying intercept and the time-varying loadings induced by the no-arbitrage restrictions.  With only models satisfying \textit{ex-ante} the no-arbitrage restrictions, we can substantially reduce the set of possible models within our model selection procedure. We derive an upper bound, and show that the number of possible models without grouping is divided by $2^3$, at least, and often by a much larger number in empirical applications.
As an example, for the model specifications with four factors used in Section~\ref{sec:empirical_result}, the set of possible models satisfying \textit{ex-ante} the no-arbitrage restrictions is $2^{97}$ times smaller than the set of possible models  without grouping. We exemplify this reduction with a simple two-factor example in Section \ref{sec:firstpass}. It echoes the discussion in \cite{giannone2021economic} that, if a prediction model with many predictors ``lacks any additional structure, then there is no hope of recovering useful information about the [high-dimensional parameter] vector with limited samples'' \cite [p.~290]{hastie2015statistical}. Imposing some constraints, hopefully driven by economic reasoning (as we promote here), should help to extract relevant information in big data problems.  As a consequence, the aOGL approach yields better performance in terms of covariate selection and estimated models without arbitrage (see our Monte Carlo results in Section \ref{sec:MC} and our empirical results in Section \ref{sec:empirical_result}).
 On our data for US single stocks,
 more than half of the stocks require dynamics in their factor loadings, while penalization without (with) grouping yields to 100\% (0\%) of all estimated time-varying models violating the no-arbitrage restrictions. Besides, the aOGL approach yields better in-sample and out-of-sample predictive performance on  an equally-weighted portfolio (see Sections \ref{sec:MC} and \ref{sec:empirical_result}). On our data for US single stocks, prediction errors are located closer to zero and their scale is narrower.
 

LASSO type techniques have already been applied successfully to factor models in finance. \cite{bryzgalova2015spurious} develops a shinkrage-based
estimator that identifies the weak factors (i.e., factors that do not correlate with the assets) and ensures consistent
and normality of the estimates of the risk premia. \cite{feng2020taming} propose a model-selection method
to evaluate the risk prices of observable factors. \cite{freyberger2020dissecting} propose a nonparametric method
to determine which firm characteristics provide incremental information for the cross section of expected
excess returns. \cite{gu2020empirical} use penalisation techniques for prediction purposes.  Alternatively, \cite{Fan2022} develop a nonparametric methodology for estimating conditional asset pricing models using deep neural networks, by employing time-varying conditional information on alphas and betas carried by firm-specific characteristics. \cite{ACMV2022} propose a novel Bayesian approach to study time-series and cross-sectional effects in asset returns, when the true factor model and its underlying parameters are uncertain. They use macro predictors to model time-variation in the factor loadings and investigate potential mispricing. While their prior beliefs are weighted against mispricing, their analysis shows that time-varying mispricing appears with a large probability. \cite{CPZ2022} use deep neural networks to estimate a stochastic discount factor model for individual stock returns and  exploit the fundamental no-arbitrage condition as criterion function, to construct the most informative test assets with an adversarial approach. Finally, let us mention that there is also work on inference for large dimensional models with observable and unobservable
factors with high frequency data \citep{fan2016incorporating,ait2017using, pelger2019state, ait2020inference}.

The outline of this paper is as follows. Section \ref{sec_model} describes the conditional linear factor models with sparse time-varying coefficients, and how to implement the no-arbitrage restrictions in the specification of the random coefficient panel model. Section~\ref{sec_estim} develops 
 our penalized two-pass regression with time-varying factor loadings. The penalization in the first-pass (time-series) regressions of Section~\ref{sec:firstpass} enforces sparsity for the time-variation drivers while also maintaining compatibility \textit{ex-ante} with the no-arbitrage restrictions through building appropriate groups of coefficients. 
 We explain in detail in Section \ref{sec:firstpass} why we prefer the aOGL method over the original Group-LASSO of \cite{yuan2006model} for the first-pass regression. The second-pass (cross-sectional) regression of Section~\ref{sec:secondpass} delivers risk premia estimates to predict equity excess returns.
 In Section~\ref{sec:secondpass}, we  show asymptotic consistency of our penalised two-pass regression  estimates under an adaptive estimation for the first-pass regression coefficients. Section \ref{sec:MC} reports our simulations results. Section~\ref{sec:empirical_result} gathers our empirical results. After describing our data on US single stocks in Section~\ref{sec_data}, we present our empirical results on  in-sample and out-of-sample prediction performances 
 in Sections \ref{sec_perf_in} and \ref{sec_perf_out}. We investigate 13 characteristics and 6 common instruments  for the dynamics of factor loadings, and use the  four-factor model of \cite{carhart1997persistence} and the five-factor model of \cite{fama2015five}. Section~\ref{sec_conclusion} concludes. We list regularity conditions in Appendix~\ref{append_regularity}, the proofs of our theoretical results in Appendices \ref{proof_lemma_asynorm_beta} and \ref{proof_theo_consitency_nu}, and a description on how to construct groups for the numerical optimisation in Appendix~\ref{append_group}.

\section{Model specification}
\label{sec_model}
In this section, we consider a conditional linear factor model with time-varying coefficients as in \citetalias{gagliardini2016time} (see \cite{gagliardini2019estimation} for a review). From their Assumptions APR.1, APR.2, and APR.3, the time-varying factor model for assets belonging to the continuum of assets $\gamma \in [0,1]$ is 

\begin{equation}
	R_{t}(\gamma)=a_{t}(\gamma)+b_{t}(\gamma)^\top f_{t}+\varepsilon_{t}(\gamma),
	\label{eq_condLinearFactMod}
\end{equation}
where $R_{t} (\gamma)$  denotes the excess return on asset $\gamma$ at period $1, \ldots, T$, vector $f_{t} \in \mathbb{R}^K$ gathers the values of the factors at date $t$. From Assumption APR.1 of \citetalias{gagliardini2016time}, the intercept $a_{t}(\gamma) \in \real$ and factor loadings
$b_{t}(\gamma) \in \real^K$ are $\mathcal{F}_{t-1}$-measurable, where the filtration process $\mathcal{F}_{t-1}$ is the information available to all investors at time $t-1$. The error terms have mean zero 
$\mathbb{E}[\varepsilon_{t}(\gamma) \vert \mathcal{F}_{t-1}] = 0$  and are uncorrelated
with the factors conditionally on information $\mathcal{F}_{t-1}$, ${\cov(\varepsilon_{t}(\gamma), f_{t,k} \vert \mathcal{F}_{t-1})} = 0$, $k=1,...,K$. Assumption APR.2 of \citetalias{gagliardini2016time}  gathers standard measurability conditions for a stochastic process, and requires that the process $\beta_t(\gamma) = (a_{t}(\gamma),b_{t}(\gamma)^\top)^{\top} \in \real^{K+1}$
is a bounded aggregate process as defined in \cite{AlNajjar_1995}, as well as  the nondegeneracy in the factor loadings across assets. Assumption APR.3 of
\citetalias{gagliardini2016time} imposes an approximate factor structure  in (\ref{eq_condLinearFactMod}) such that, for any sequence $\gamma_i \in [0,1], i = 1, \ldots, n$, with $\Sigma_{\varepsilon_{t},t,n} \in \real^{n\times n}$ being the conditional variance-covariance matrix of the vector $(\varepsilon_{t}(\gamma_1), \ldots ,\varepsilon_{t}(\gamma_n))^{\top}$ knowing $Z_{t-1}$, there exists a set such that $n^{-1}\eig_{\text{max}}(\Sigma_{\varepsilon_{t},t,n} ) \overset{\text{\tiny{$L^2$}}}{\longrightarrow}0$ as $n\to\infty$, where $\eig_{\text{max}}(\Sigma_{\varepsilon_{t},t,n})$ denotes the largest eigenvalue of $\Sigma_{\varepsilon_{t},t,n}$, and where $\overset{\text{\tiny{$L^2$}}}{\longrightarrow}$ denotes convergence in the $L^2$-norm. Under Assumptions APR.4 of  \citetalias{gagliardini2016time}, the following asset pricing restriction holds:
\begin{equation}
	a_{t}(\gamma)=b_{t}(\gamma)^{\top}\nu_{t},
	\label{eq_condNoArbRest}
\end{equation}
 for all $\gamma \in [0,1]$, at any date $t = 1, 2, \ldots$ where random vector $\nu_{t}\in\mathbb{R}^{K}$ is unique and is $\mathcal{F}_{t-1}$-measurable, which can also be written as 
\begin{equation}
    \mathbb{E}\left[R_{t}(\gamma)\vert\mathcal{F}_{t-1}\right]=b_{t}(\gamma)^{\top}\lambda_{t},
    \label{eq_expect}
\end{equation}
 with $\lambda_{t}=\nu_{t}+\mathbb{E}[f_{t}\vert\mathcal{F}_{t-1}] \in \real^K$. Equation (\ref{eq_expect}) shows the link between expected excess returns and the product of the time-varying factor loadings and risk premia. Below, we rely  on that link to predict excess returns. Assumption APR.4 of \citetalias{gagliardini2016time} excludes asymptotic arbitrage opportunity, such that there is no portfolio sequence with zero cost and positive payoff.
The conditioning information $\mathcal{F}_{t-1}$
contains $Z_{\underline{t-1}}$ and $Z_{\underline{t-1}}(\gamma)$, where $Z_{t-1}\in\mathbb{R}^{p}$ is  a vector of lagged instruments common to all stocks, $Z_{t-1}(\gamma)\in\mathbb{R}^{q}$, for $\gamma \in [0,1]$, is a vector of
lagged characteristics  specific
to stock $\gamma$, and  $Z_{\underline{t}}=\{Z_{t},Z_{t-1},...\}$ denotes the set of past realizations.
Vector $Z_{t-1}$ may include past observations of
the factors and some additional variables such as macroeconomic variables.
Vector $Z_{t-1}(\gamma)$ may include past observations of firm characteristics and stock returns.  We define the dynamics of the factor loadings $b_t(\gamma)$ as a sparse linear function of $Z_{t-1}$ \citep{Shanken_1990,Ferson_Harvey_1991} and $Z_{t-1}(\gamma)$ \citep{Avramov_Chordia_2006}.

\begin{Aassumption} (Sparse time-varying factor loadings) \\
    The factor loadings are such that $b_{t}(\gamma) = A(\gamma) + B(\gamma) Z_{t-1} + C(\gamma) Z_{t-1}(\gamma)$, where  $A(\gamma)\in \real^{K}$ correspond to a time-invariant model, and $B(\gamma) \in \real^{K \times p}$, $C(\gamma)\in \real^{K \times q}$ are sparse matrices of coefficient for any $\gamma \in [0,1]$ and any $t$.
    \label{assum_spec_fact_load}
\end{Aassumption}
Moreover, we define the vector of risk premia as a sparse linear function of lagged instruments $Z_{t-1}$ \citep{Cochrane_1996,Jagannathan_Wang_1996} and specify the conditional expectation of the factor $\mathbb{E}\left[ f_t \vert \mathcal{F}_{t-1} \right]$ given the filtration process $\mathcal{F}_{t-1}$.
\begin{Aassumption}(Sparse time-varying risk premia) \\
    The risk premia vector is such that 
    
    (i) $\lambda_t = \Lambda_0 + \Lambda_1 Z_{t-1}$, where $\Lambda_0\in \real^{K}$ correspond to a time-invariant model and $\Lambda_1 \in \real^{K \times p}$ is a sparse matrix  for any $t$. 
    
    \noindent The conditional expectation of the factor is such that
    
    (ii) $\mathbb{E}\left[ f_t \vert \mathcal{F}_{t-1} \right] = F_0 + F_1 Z_{t-1}$, where $F_0 \in \real^{K}$ corresponds to a time-invariant model and   $F_1 \in \real^{K \times p}$ is a sparse matrix for any $t$.
    \label{assum_spec_riks_premia}
\end{Aassumption}
Assumptions~\ref{assum_spec_fact_load} and \ref{assum_spec_riks_premia} differ from Assumptions FS.1 and FS.2 of \citetalias{gagliardini2016time}. Indeed, we consider here the matrices $B(\gamma), C(\gamma), \Lambda_1$ and $F_1$ of coefficients as sparse, meaning that only a small fraction of the $Z_{t-1}$ or $Z_{t-1}(\gamma)$ for $\gamma \in [0,1]$ are useful to describe the dynamics of the factor loadings, risk premia, and conditional expectation of the factors. 
Building on the sampling scheme from Assumptions SC.1 and SC.2 of \citetalias{gagliardini2016time}, we define the indicator variable $I_t(\gamma)$, for all $\gamma \in [0,1]$, such that $I_t(\gamma) = 1$ if the return on asset $\gamma$ is observable at time $t$, and 0 if not. Assumption SC.1 ensures that $I_t(\gamma)$, $\varepsilon_t(\gamma)$ and variables in $\mathcal{F}_{t-1}$ are independent, while Assumption SC.2  ensures that the random variables $\gamma_{i}$,
$i=1,...,n$, are i.i.d.\ indices, independent of $\varepsilon_{t}(\gamma)$,
$I_{t}(\gamma)$, and $\mathcal{F}_{t-1}$. From the above sampling scheme, we can now use the following notation: $I_{i,t} = I_t(\gamma_i), R_{i,t} = R_t(\gamma_i),\beta_{i,t} = \beta_t(\gamma_i), \varepsilon_{i,t} = \varepsilon_{t}(\gamma_i),  A_i = A(\gamma_i), B_i = B(\gamma_i), C_i = C(\gamma_i)$ and $Z_{i,t-1} = Z_{t-1}(\gamma_i) $ as well as $a_{i,t} = a_t(\gamma_i)$ and $b_{i,t} = b_t(\gamma_i)$. Hence, from Assumptions~\ref{assum_spec_fact_load} and \ref{assum_spec_riks_premia}, we can express  (\ref{eq_condLinearFactMod}) using the asset pricing restriction in (\ref{eq_condNoArbRest})  as the following Data Generating Process (DGP):
\begin{equation}
	\begin{aligned}
		R_{i,t} &= A_i^{\top} \left(\Lambda_0 - F_0\right) + A_i^{\top} \left(\Lambda_1 - F_1\right) Z_{t-1} + Z_{t-1}^{\top} B_i^{\top} \left(\Lambda_0 - F_0\right)  \\ 
		&+ Z_{t-1}^{\top} B_i^{\top} \left(\Lambda_1 - F_1\right)Z_{t-1} + Z_{i,t-1}^{\top} C_i^{\top} \left(\Lambda_0 - F_0\right) \\ &+  Z_{i,t-1}^{\top} C_i^{\top} \left(\Lambda_1 - F_1\right)Z_{t-1}	+ A_i^{\top}f_{t} +  Z_{t-1}^{\top} B_i^{\top}f_{t} + Z_{i,t-1}^{\top} C_i^{\top} f_{t}  + \varepsilon_{i,t}.
	\end{aligned}
	\label{eq_dgp}
\end{equation}
We see that the first term $A_i^{\top} \left(\Lambda_0 - F_0\right)$ corresponds to the time-invariant part in the time-varying intercept $a_{i,t}$, while the term $A_i^{\top}f_{t}$ corresponds to the time-invariant part of the time-varying  factor loadings $b_{i,t}$. To separate the time-invariant part from the time-varying part, we make the following assumption on the model specification.

\begin{Aassumption} (Non sparse time-invariant contribution) \\
\label{assum_time_ivariant}
    We define the time-invariant contribution as $            A_i^{\top} \left(\Lambda_0 - F_0\right) + A_i^{\top}f_{t}.$ We require that 
       the vectors $A_i \in \real^K, \Lambda_0 \in \real^{K}$, and $F_0 \in \real^{K}$ have a full vector specification, i.e., do not contain null-elements.
\end{Aassumption}
Assumption \ref{assum_time_ivariant} ensures that the time-invariant part of a factor loading is always included in the model specification, so that we can distinguish a factor with a time-invariant loading from a factor 
with a time-varying loading for asset $i$. This assumption is key to analyze which instrument $Z_{t-1}$ and  characteristic $Z_{i,t-1}$, if needed, drive the dynamics of the factor loadings $b_{i,t}$ for asset $i$, and impact on the prediction  $\mathbb{E}[R_{i,t}\vert\mathcal{F}_{t-1}]$ via  (\ref{eq_expect}). 
Since implementing a penalized two-pass regression given on (\ref{eq_dgp}) is difficult (due to the quadratic form in lagged instruments $Z_{t-1}$ and  $Z_{i,t-1}$), we redefine the regressors and coefficients, as a generic  panel model. Beforehand,  let us define the vector of lagged instruments including the intercept as $\tilde{Z}_{t-1} = (1, Z_{t-1}^{\top})^{\top} \in \real^{\tilde{p}}$, where $\tilde{p} = p+1$, and the new matrices $\breve{B_i} = [A_i | B_i ] \in \real^{K \times \tilde{p}} $ and ${\Lambda} - {F} = [(\Lambda_0 - F_0) | (\Lambda_1 - F_1) ] \in \real^{K \times \tilde{p}} $ that stack respectively column-wise the elements of $A_i$, $B_i$, and $(\Lambda_0 - F_0) ,(\Lambda_1 - F_1 )$. The linear transformed regressors are
\begin{equation*}
    x_{2,i,t} = \left(x_{21,i,t}^{\top}, x_{22,i,t}^{\top}\right)^{\top} =\left( f_{t}^{\top}\otimes \tilde{Z}_{t-1}^{\top} , f_{t}^{\top}\otimes  Z_{i,t-1}^{\top}\right)^{\top}\in\mathbb{R}^{d_{2}},
\end{equation*}
where $d_2 = d_{21} + d_{22} =  K\tilde{p} + Kq$, and 
\begin{equation*}
    x_{1,i,t} = \left(x_{11,i,t}^{\top}, x_{12,i,t}^{\top}\right)^{\top}=\left(\vech\left[X_{t}\right]^{\top},\tilde{Z}_{t-1}^{\top}\otimes Z_{i,t-1}^{\top}\right)^{\top}\in\mathbb{R}^{d_{1}},
\end{equation*}
where  $d_{1} = d_{11} + d_{12} =(\tilde{p}+1)\tilde{p}/2+\tilde{p}q$ and the symmetric matrix $X_{t}=(X_{t,k,l})_{k,l}\in\mathbb{R}^{\tilde{p}\times \tilde{p}}$ is such that $X_{t,k,l}=\tilde{Z}_{t-1,k}^{2}$, if $k=l$, and $X_{t,k,l}=2\tilde{Z}_{t-1,k}\tilde{Z}_{t-1,l}$,
otherwise, for $k,l=1,\ldots,\tilde{p}$, where $\tilde{Z}_{t,k}$ denotes the $k$-th
component of the vector $\tilde{Z}_{t}$. The vector-half operator $\vech\left[\cdot\right]$
stacks the elements of the lower triangular part of a $\tilde{p}\times \tilde{p}$
matrix as a $\tilde{p}\left(\tilde{p}+1\right)/2 $ vector.  The first element of $\vech\left(X_{t}\right)$ is related to the time-invariant coefficients $A_i^{\top} \left(\Lambda_0 - F_0\right)$, whereas the elements $2, \ldots, \tilde{p}$ are related to $A_i^{\top} \left(\Lambda_1 - F_1\right) Z_{t-1} + Z_{t-1}^{\top} B_i^{\top} \left(\Lambda_0 - F_0\right)$. Through the above redefinitions of the regressor, we can write (\ref{eq_dgp}) as 
\begin{equation}
	R_{i,t}=\beta_{i}^{\top}x_{i,t}+\varepsilon_{i,t},
	\label{eq_LinearFactorRegression}
\end{equation}
where $x_{i,t} = (x_{1,i,t}^{\top}, x_{2,i,t}^{\top})^{\top}$ is of dimension $d = d_1 + d_2$ and $\beta_i = (\beta_{1,i}^{\top}, \beta_{2,i}^{\top})^{\top}$ is defined as 
\begin{equation}
    \begin{aligned}
         \betaone &= \left( \beta_{11,i}^{\top} , \beta_{12,i}^{\top}\right)^{\top} \in \real^{d_1}, \\ 
         \beta_{11,i} &= N_{\tilde{p}} \left[\left({\Lambda} - {F}\right)^{\top}\otimes I_{\tilde{p}}\right] \vect[\breve{B}_i^{\top}] \in \real^{d_{11}} ,\\
         \beta_{12,i} &= W_{\tilde{p},q}\left[\left({\Lambda} - {F}\right)^{\top}\otimes I_q\right] \vect[C_i^{\top}] \in \real^{d_{12}} ,\\
         N_{\tilde{p}} &=\frac{1}{2}D_{\tilde{p}}^{+}(W_{\tilde{p}}+I_{\tilde{p}^{2}}) \in \real^{[(\tilde{p}+1)\tilde{p}/2+\tilde{p}q] \times \tilde{p}^2} , \\ 
         \betatwo &= \left({\beta}_{21,i}^{\top},{\beta}_{22,i}^{\top}\right)^{\top} =  \left(\vect[\breve{B}_i^{\top}]^{\top},\vect[C_i^{\top}]^{\top}\right)^{\top} \in \real^{d_2} ,
    \end{aligned}
    \label{betaone}
\end{equation}
and where $W_{\tilde{p},q}$ is the commutation matrix such that $\vect[M^{\top}] = W_{\tilde{p},q} \vect[M]$. Moreover, $D_{\tilde{p}}^{+}$ denotes the $((\tilde{p}+1)\tilde{p}/2+\tilde{p}q) \times \tilde{p}^2$ Moore-Penrose inverse of the duplication matrix $D_{\tilde{p}}$ such that $\vech[M] = D_{\tilde{p}}^{+} \vect[M]$, for any matrix $\tilde{p} \times \tilde{p}$ matrix $M$. 
%
%
The following section describes the selection and estimation part of the model.

\section{Estimation and selection}
\label{sec_estim} 

This section implements the two-pass regression of \cite{jensen1972capital} and \cite{fama1973risk},  while selecting the contributing variables in the time-varying factor loadings.  The penalized first-pass (time-series) regression selects and estimates the non-zero coefficients  $\beta_i$ for $i = 1, \ldots, n$, ensuring a model specification compatible \textit{ex-ante} with the no-arbitrage restrictions through the aOGL approach of \cite{percival2012theoretical}. The second-pass regression relies on the Weighted Least-Square (WLS) estimator of \citetalias{gagliardini2016time} to estimate the vector $\nu$, and takes the adaptive LASSO (aLASSO) estimator of \cite{zou2006adaptive} to select and estimate the matrix $F$ of coefficients of the conditional expectation of the factors.

\subsection{First-pass regression}
\label{sec:firstpass}

The goal of the penalized first-pass regression is to select and estimate the factor loadings for each asset $i = 1, \ldots, n$, while keeping their respective time-invariant contribution fully specified as described in Assumption~\ref{assum_time_ivariant}. Moreover, it aims at selecting variables ensuring a proper model specification consistent \textit{ex-ante} with the no-arbitrage restrictions for each stock. A possible solution to ensure that these restrictions are satisfied while allowing to select variables in the first-pass regression is to consider a LASSO-type estimator based on appropriate predefined sets of indices corresponding to groups of variables. We define $\groupset \subset \mathcal{P}(\{1, \ldots, d\})$ as the set of indices corresponding to all possible (potentially overlapping) groups in line with the no-arbitrage restrictions, where $\mathcal{P}(\{1, \ldots, d\})$ denotes the power set of $\{1, \ldots, d\}$. Moreover, we let $g \in \groupset$ denote a possible group and we require that the indices associated to all covariates belong to at least one group. Under the framework discussed in the previous sections, we define below the restrictions on $\groupset$ such that a model selection procedure based on $\groupset$ satisfies \textit{ex-ante}  the no-arbitrage restrictions  by construction.
\setcounter{Restriction}{0}
\renewcommand{\theHRestriction}{otherRestriction\theRestriction}
\renewcommand\theRestriction{\Alph{Restriction}}
\begin{RRestriction}
    \label{res:ti}
    The time-invariant coefficients belong to a single group, where no amount of shrinkage is applied.
\end{RRestriction}

\begin{RRestriction}
    \label{res:no_group}
    Each covariate related to the non-diagonal elements of $X_t$ belongs to a single group.
\end{RRestriction}

\begin{RRestriction}
    \label{res:zt}
    For instrument $\tilde{Z}_{t-1,l}$, for $l= 1, \ldots, \tilde{p}$,  if all its  corresponding $\tilde{Z}_{t-1,l}f_{t,k}$,  for $k = 1, \ldots, K$, in $x_{2,i,t}$ are not included in the estimated model, only the regressors $\tilde{Z}_{t-1,l}^2$, related to the diagonal element of $X_t$, in $x_{1,i,t}$   should not be included. For  characteristic $Z_{i,t-1,m}$, for $m=1, \ldots, q$, if all its  corresponding  $Z_{i,t-1,m}f_{t,k}$ for $k = 1, \ldots, K$, in $x_{2,i,t}$ are not included in the estimated model, only the regressors  $Z_{i,t-1,m}$ in $x_{1,i,t}$   should not be included.
\end{RRestriction}

\begin{RRestriction}
    \label{res:zi}
    For instrument $\tilde{Z}_{t-1,l}$, for $l= 1, \ldots, \tilde{p}$, if at least one of its corresponding  $\tilde{Z}_{t-1,l}f_{t,k}$, for $k = 1, \ldots, K$, in $x_{2,i,t}$ are included in the estimated model, only the regressors $\tilde{Z}_{t-1,l}^2$, related to the diagonal element of $X_t$, in $x_{1,i,t}$   should be included.
     For characteristic $Z_{i,t-1,m}$, for $m=1, \ldots, q$, if at least one of its  corresponding $Z_{i,t-1,m}f_{t,k}$, for $k = 1, \ldots, K$, in $x_{2,i,t}$ are included in the estimated model, only the regressors  $Z_{i,t-1,m}$ in $x_{1,i,t}$   should be included.
\end{RRestriction}
These restrictions ensure that Assumption~\ref{assum_time_ivariant} is satisfied and that a model selection procedure guarantees that the instrument $\tilde{Z}_{t-1,l}$ or characteristic $Z_{i,t-1,m}$ exist in either both $x_{1,i,t}$ and $x_{2,i,t}$, or neither. More specifically, Restriction~\ref{res:ti} is related to Assumption~\ref{assum_time_ivariant}, which requires the coefficients in $\beta_i$ related to the time-invariant contribution to be always included in the selected model. Restriction~\ref{res:no_group} is related to  Assumption~\ref{assum_spec_fact_load} and Assumption~\ref{assum_spec_riks_premia}.  Under the DGP in (\ref{eq_dgp}), and from the definition of $\vech(X_t)$, we can see that the off-diagonal of $X_t$ in $\vech[X_t]$ cannot be assigned to any groups. We cannot assign $2\tilde{Z}_{t-1,s}\tilde{Z}_{t-1,l}$ to a group a priori, since its contribution can come from either the specification in Assumption~\ref{assum_spec_fact_load} or \ref{assum_spec_riks_premia}. Restriction~\ref{res:no_group} reflects this point, and imposes no specific group-structure to those covariates which are penalized individually. Restrictions~\ref{res:zt} and \ref{res:zi} are critical in the model building. They constrain the set of possible models only to those compatible with the no-arbitrage restrictions, so that we do not introduce arbitrage \textit{ex-ante} in the model specified in (\ref{eq_LinearFactorRegression}).
We want to avoid that the no-arbitrage restriction $a_{i,t} = b_{i,t}^{\top} \nu_t$ is violated by construction \textit{ex-ante} in the specification. 

To satisfy the above restrictions, the Group-LASSO of \cite{yuan2006model} constrains the set of possible models. For its implementation, we need to create a group with all scaled factors and their corresponding terms in the intercept, hence it implies that we select  either all scaled factors (keep the group) or none of them (delete the group). To illlustrate this point, let us consider the following simple case with one common instrument, say inflation, and the Fama-French five-factor model \citep{fama2015five}. The Group-LASSO would force us to select either all scaled factors (product between lagged inflation and the factors), or none of them. It removes the possibility that only a subset of them is relevant; for example, only the product of inflation and the market factor matters for the dynamics of excess returns. Besides, we could think of using multiple groups, each one containing one scaled factor and its associated instrument. \cite{jacob2009group} investigate such a proposal and show that this approach is not appropriate as the Group-LASSO removes all groups if at least one of those groups is not selected.

To tackle this problem, \cite{jacob2009group} propose the OGL, or latent Group-LASSO. They introduce the latent variables ${v}_{\group} \in \mathcal{V}_g = \{ x \in \real^d | \supp(x) = g \}$, for $g \in \groupset$ and where $\supp(x)$ denotes the support of $x$, i.e., the set of indices $i \in \{1, \ldots, d\}$ such that $x_i \neq 0$. Moreover, we define ${v_g} = (v_{g_1}^\top, \ldots, v_{g_{J}}^\top)^\top \in \real^d, \mathcal{V}(\beta) = \{v_g : g\in \groupset\}$, s.t.\ $\beta = \sum_{g\in \groupset} {v}_{g}$, and $J = |\groupset|$, $|\cdot|$ denotes the cardinality of a set and $g_j, \, j = 1,\ldots, J$, denotes the $j$-th element of $\groupset$. 
 Hence, the OGL estimator is the solution of the following optimization problem: 
\begin{equation}
    \hat{\beta}_{i} = \argmin_{\beta_i \in \real^d} \;  \frac{1}{T_i}\sum_{t} \left(I_{i,t} R_{i,t} - \beta_i^{\top} I_{i,t}x_{i,t}\right)^2 + 2\delta \|\beta_i\|_{2,1,\groupset}, 
    \label{eq_ogl}
\end{equation}
with the penalty term $\|\beta_i\|_{2,1,\groupset}$ defined as
\begin{equation}
    \|\beta_i\|_{2,1,\groupset} =   \min_{\mathcal{V}(\beta)} \sum_{g \in \groupset} \|{v}_g\|,
    \label{eq_norm_ball}
\end{equation}
where $\|\cdot\|$ denotes the $l_2$-norm. In this work, we consider the adaptive version of OGL (aOGL) studied by \cite{percival2012theoretical}, for which the estimator is described as follow: 
\begin{equation}
    \hat{\beta}_{i} = \argmin_{\beta_i \in \real^d} \;  \frac{1}{T_i}\sum_{t} \left(I_{i,t} R_{i,t} - \beta_i^{\top} I_{i,t}x_{i,t}\right)^2 + 2\delta \min_{\mathcal{V}(\beta)} \delta_g \sum_{g \in \groupset} \|{v}_g\|,
    \label{eq_aogl}
\end{equation}
where $\delta_{\group} \geq 0$ denotes the data-dependent (adaptive) weight  associated to group $g$, and $\delta \geq 0$ corresponds to the overall amount of shrinkage. There are different strategies available in the literature for the Group LASSO and OGL to get estimator consistency and support selection consistency. They are based on the irrepresentable condition \citep{bach2008consistency,jacob2009group}, adaptive shrinkage \citep{nardi2008asymptotic,percival2012theoretical} and group sparsity \citep{lounici2011oracle}.  We choose adaptive shrinkage since it simplifies the presentation and derivation of our asymptotic results in a random design setting.  Since our goal is to shrink toward the model that includes only the time-invariant contribution of the covariates, the weight associated with the first element of $\delta_{\group}$ is equal to zero. The penalty term in (\ref{eq_norm_ball}) leads to a solution which is a union of the groups due to the latent variables ${v}_{g}$. One strategy to solve the minimization problem given in (\ref{eq_ogl}) and (\ref{eq_aogl}) is the  duplication of covariates  put forward in \cite{jacob2009group}, that we adapt to our setting. In line with Restrictions~\ref{res:ti} to \ref{res:zi}, we consider 4 different group types. The first group includes the time-invariant intercept and time-invariant factors, and is not penalised. The second set of groups contains the covariates related to Restriction~\ref{res:no_group}, which are penalized individually. The next two sets of groups consider Restrictions~\ref{res:zt} and \ref{res:zi}. They respectively group the terms in $Z^2_{t-1}$ and $Z_{i,t-1}$ from  $x_{1,i,t}$ with their corresponding scaled factors in $x_{2,i,t}$. The columns of the initial vector with the elements indexed by the group $g$, which need to be duplicated, create a new vector of duplicated regressors. Then, we can solve the optimization problem in (\ref{eq_aogl}) considering the duplicated regressors (instead of the initial ones), using the existing standard algorithm for the Group-LASSO.  Appendix~\ref{append_group} describes in detail how to construct those groups complying with the no-arbitrage restrictions \textit{ex-ante}, and yielding the full vector of duplicated regressors used in the numerical optimisation.

 Let us now compare the number of possible models under aOGL and aLASSO methods. For the aOGL approach, we can associate a model to every subset of $\mathcal{G}$. Indeed, consider $\mathcal{W}  \subseteq \groupset$, then this subset is associated to the  set $S_{\mathcal{W}} = \bigcup_{l = 1}^{|\mathcal{W}|} \mathcal{W}_l$ of indices. It allows us to enumerate the number $2^{J-1}$ of possible models under appropriate grouping. That number is typically much lower in empirical applications than the number $2^{d-n_1}$ of possible models with a LASSO penalization, where $n_1$ is the number of covariates associated to the time-invariant contribution group. 
 We get the ratio 
 $2^{J-1}/2^{d-n_1} = 2^{-(pq+p+q)}$, and we can see that, for large $p$ and $q$, the aLASSO method examines  many more possibilities.
 Besides, from Assumption~\ref{assum_spec_fact_load}, we have $\min(p,q) \geq 1$,  and deduce the upper bound:
 \begin{equation}
 \frac{2^{J-1}}{2^{d - n_1}} \leq \frac{1}{8}. \label{upper_bound}
\end{equation}

To further illustrate the grouping structure and the importance of Restrictions~\ref{res:ti} to \ref{res:zi}, let us consider the following simple two-factor model with a single common instrument and a single characteristic. Here, we have $K=2$, $\tilde{p} = 2$, and $q = 1$, with $\tilde{Z}_{t-1} = (1,Z_{t-1})^{\top} \in \real^2$, so that the regressors $x_{i,t} = (x_{1,i,t}^{\top},x_{2,i,t}^{\top})^{\top}$ become
\begin{equation*}
    \begin{aligned}
    x_{1,i,t} &= (x_{1,i,t,1},x_{1,i,t,2},x_{1,i,t,3},x_{1,i,t,4},x_{1,i,t,5})^{\top} \\
    &=  (1, 2Z_{t-1}, Z^2_{t-1}, Z_{i,t-1}, Z_{t-1} Z_{i,t-1})^{\top} \in \real^5,
    \end{aligned}
\end{equation*}
and 
\begin{equation*}
    \begin{aligned}
    x_{2,i,t} &= (x_{2,i,t,1},x_{2,i,t,2},x_{2,i,t,3},x_{2,i,t,4},x_{2,i,t,5},x_{2,i,t,6})^{\top} \\
    &= (f_{t,1},Z_{t-1}f_{t,1},f_{t,2},Z_{t-1}f_{t,2},Z_{i,t-1}f_{t,1},Z_{i,t-1}f_{t,2})^{\top} \in \real^6,
    \end{aligned}
\end{equation*}
with their respective coefficients $\beta_{1,i} = (\beta_{1,i,1},\beta_{1,i,2},\beta_{1,i,3},\beta_{1,i,4},\beta_{1,i,5})^{\top}$ and $\beta_{2,i} = (\beta_{2,i,1},\beta_{2,i,2},\beta_{2,i,3},\beta_{2,i,4},\beta_{2,i,5},\beta_{2,i,6})^{\top}$. 
\begin{table}
\footnotesize
	\centering
	\begin{tabular}{cccccccccccc}
	\toprule
	\centering   & $x_{1,1}$ & $x_{1,2}$ & $x_{1,3}$ & $x_{1,4}$ & $x_{1,5}$ & $x_{2,1}$ & $x_{2,2}$ & $x_{2,3}$ & $x_{2,4}$ & $x_{2,5}$ & $x_{2,6}$\\
	 \cmidrule{2-12}
	$\model_1$ & \possible & \notpossible & \notpossible & \notpossible & \notpossible &\possible & \notpossible &\possible & \notpossible & \notpossible & \notpossible \\ 
	 \cmidrule{2-12}
	$\model_2$ & \possible & \possible & \notpossible & \notpossible & \notpossible &\possible & \notpossible &\possible & \notpossible & \notpossible & \notpossible \\ 
	\cmidrule{2-12}
	$\model_3$ & \possible & \notpossible & \possible & \notpossible & \notpossible &\possible & \possible &\possible & \notpossible & \notpossible & \notpossible \\ 
	\cmidrule{2-12}
	$\model_4$ & \possible & \notpossible & \possible & \notpossible & \notpossible &\possible & \notpossible &\possible & \possible & \notpossible & \notpossible \\ 
	\cmidrule{2-12}
	$\model_5$ & \possible & \possible & \possible & \notpossible & \notpossible &\possible & \possible &\possible & \notpossible & \notpossible & \notpossible \\ 
	\cmidrule{2-12}
	$\model_6$ & \possible & \possible & \possible & \notpossible & \notpossible &\possible &  \notpossible & \possible & \possible & \notpossible &  \notpossible \\
	\cmidrule{2-12}
	$\model_7$ & \possible & \notpossible & \possible & \notpossible & \notpossible &\possible & \possible & \possible & \possible & \notpossible & \notpossible \\ 
	\cmidrule{2-12}
	$\model_8$ & \possible & \possible & \possible & \notpossible & \notpossible &\possible & \possible & \possible & \possible & \notpossible & \notpossible \\
	\cmidrule{2-12}
	$\model_9$ & \possible & \notpossible & \notpossible & \possible & \possible &\possible & \notpossible &\possible & \notpossible & \possible &  \notpossible \\ 
	\cmidrule{2-12}
	$\model_{10}$ & \possible & \notpossible & \notpossible & \possible & \possible &\possible & \notpossible & \possible & \notpossible & \notpossible & \possible \\ 
	\cmidrule{2-12}
	$\model_{11}$ & \possible & \notpossible & \notpossible & \possible & \possible &\possible & \notpossible &\possible & \notpossible & \possible & \possible \\ 
	\cmidrule{2-12}
	$\model_{12}$ & \possible & \possible & \notpossible & \possible & \possible &\possible & \notpossible &\possible & \notpossible & \possible & \notpossible \\ 
	\cmidrule{2-12}
	$\model_{13}$ & \possible & \possible & \notpossible & \possible & \possible &\possible & \notpossible &\possible & \notpossible & \notpossible & \possible \\ 
	\cmidrule{2-12}
	$\model_{14}$ & \possible & \possible & \notpossible & \possible & \possible &\possible & \notpossible &\possible & \notpossible & \possible & \possible \\ 
	\cmidrule{2-12}
	$\model_{15}$ & \possible & \notpossible & \possible & \possible & \possible &\possible & \possible &\possible & \notpossible & \possible & \notpossible \\ 
	\cmidrule{2-12}
	$\model_{16}$ & \possible & \notpossible & \possible & \possible & \possible &\possible & \possible &\possible & \notpossible & \notpossible & \possible \\ 
	\cmidrule{2-12}
	$\model_{17}$ & \possible & \notpossible & \possible & \possible & \possible &\possible & \possible &\possible & \notpossible & \possible & \possible \\ 
	\cmidrule{2-12}
	$\model_{18}$ & \possible & \possible & \possible & \possible & \possible & \possible & \possible &\possible & \notpossible & \possible & \notpossible \\ 
	\cmidrule{2-12}
	$\model_{19}$ & \possible & \possible & \possible & \possible & \possible &\possible & \possible &\possible & \notpossible & \notpossible & \possible \\ 
	\cmidrule{2-12}
	$\model_{20}$ & \possible & \possible & \possible & \possible & \possible &\possible & \possible &\possible & \notpossible & \possible & \possible \\ 
	\cmidrule{2-12}
	$\model_{21}$ & \possible & \notpossible & \possible & \possible & \possible & \possible &  \notpossible & \possible & \possible & \possible & \notpossible \\ 
	\cmidrule{2-12}
	$\model_{22}$ & \possible & \notpossible & \possible & \possible & \possible & \possible & \notpossible &\possible & \possible & \notpossible & \possible \\ 
	\cmidrule{2-12}
	$\model_{23}$ & \possible & \notpossible & \possible & \possible & \possible &\possible & \notpossible &\possible & \possible & \possible & \possible \\ 
	\cmidrule{2-12}
	$\model_{24}$ & \possible & \possible & \possible & \possible & \possible &\possible & \notpossible &\possible &\possible & \possible &  \notpossible \\ 
	\cmidrule{2-12}
	$\model_{25}$ & \possible & \possible & \possible & \possible & \possible &\possible & \notpossible &\possible &\possible  & \notpossible & \possible \\ 
	\cmidrule{2-12}
	$\model_{26}$ & \possible & \possible & \possible & \possible & \possible &\possible &  \notpossible &\possible & \possible & \possible & \possible \\ 
	\cmidrule{2-12}
	$\model_{27}$ & \possible & \notpossible & \possible & \possible & \possible &\possible &  \possible  &\possible & \possible & \possible &  \notpossible \\ 
	\cmidrule{2-12}
	$\model_{28}$ & \possible & \notpossible & \possible & \possible & \possible &\possible &  \possible  &\possible & \possible &  \notpossible & \possible \\ 
	\cmidrule{2-12}
	$\model_{29}$ & \possible & \notpossible & \possible & \possible & \possible &\possible &  \possible  &\possible & \possible & \possible & \possible \\ 
	\cmidrule{2-12}
	$\model_{30}$ & \possible & \possible & \possible & \possible & \possible &\possible &  \possible  &\possible &\possible & \possible &  \notpossible \\ 
	\cmidrule{2-12}
	$\model_{31}$ & \possible & \possible & \possible & \possible & \possible &\possible &  \possible  &\possible &\possible  &  \notpossible & \possible \\ 
	\cmidrule{2-12}
	$\model_{32}$ & \possible & \possible & \possible & \possible & \possible &\possible &  \possible   &\possible & \possible & \possible & \possible \\
	\bottomrule
	\end{tabular}
	\caption{Set of possible models according to Restrictions~\ref{res:ti}-\ref{res:zi} when $K=2$, $\tilde{p} = 2$, and $q = 1$. A check denotes inclusion of a covariate in model $\model_j$. A cross denotes exclusion of a covariate in  $\model_j$. For notational simplicity, we remove $i$ and $t$ in the column labeling such that $x_{l,i,t,k} = x_{l,k}$}.
    \label{tab_model_possible}
\end{table}
\sloppy From the definition of  grouping structure in Apprendix~\ref{append_group}, we construct the set of six groups made of the covariates: \break $(x_{1,i,t,1}, x_{2,i,t,1}, x_{2,i,t,3})^{\top}$ for the time-invariant contribution,  $(x_{1,i,t,2})$ for  the covariate associtated to Restriction~\ref{res:no_group},  $(x_{1,i,t,3}, x_{2,i,t,2})^{\top}$
and  $(x_{1,i,t,3}, x_{2,i,t,4})^{\top}$ grouping the covariates in $\tilde{Z}_{t-1}$,  and  finally $(x_{1,i,t,4}, x_{1,i,t,5}, x_{2,i,t,5})^{\top}$ and  $ (x_{1,i,t,4}, x_{1,i,t,5}, x_{2,i,t,6})^{\top}$
 grouping the covariates in $\tilde{Z}_{i,t-1}$. Stacking those vectors row-wise in a single column
defines the full vector of  duplicated covariates  for the numerical optimisation in the aOGL estimation. Besides, we can use this simple example to illustrate two possible manners to introduce \textit{ex-ante} arbitrage through careless modeling. Removing the covariates $x_{2,i,t,2} = Z_{t-1} f_{t,1}$ and $x_{2,i,t,4} = Z_{t-1} f_{t,2}$ from the full model might introduce \textit{ex-ante} arbitrage through $x_{1,i,t,3} = Z_{t-1}^2$ since we miss its associated scaled factors in  $x_{2,i,t}$. Here, the coefficient associated with $x_{1,i,t,3}$ might be shrunk to zero by the aLASSO estimator, avoiding \textit{ex-post} a model with arbitrage. On the contrary, removing the quadratic term $x_{1,i,t,3}$, while keeping its corresponding scaled factors $x_{2,i,t,2}$ and $x_{2,i,t,4}$, introduces \textit{ex-ante} arbitrage in the model by construction, since we cannot estimate the coefficient of $x_{1,i,t,3}$, when that covariate is absent from the model.

Table \ref{tab_model_possible}  lists the set $\model = \{\model_1, \ldots, \model_{32}\}$ of possible models that respect Restrictions~\ref{res:ti} to \ref{res:zi} with $\model_1$ being the model with the time-invariant contribution only (Assumption \ref{assum_time_ivariant}). 
The aOGL method gives $2^5$ possible models. It is 
considerably
smaller than the $2^{8} = 256$ possible models under the aLASSO method. 
Here, we reach the upper bound (\ref{upper_bound}) since $p = q = 1$. 
We can see that our regularization approach restricts the space of searched models, even in this simple time-varying setting,  and hence permits a sound exploration of the possible models consistent with finance theory. Moreover, the two specifications with arbitrage described in the above lines are not in the set $\model$
of models induced by the grouping structure of the aOGL approach, strengthening conducive arguments for our proposed method. 

 Having showed the advantages of the aOGL in terms of model building, we now state the asymptotic result of the first-pass regression.  Beforehand, we introduce some notations from \cite{percival2012theoretical}. Let us define the two sets of indices \break $H_i=\left\{l \in \{1, \ldots, d\}: \beta_{i,l} \neq 0\right\}$, $H^c_i= \{l \in \{1, \ldots, d\}: \beta_{i,l} = 0\}$, corresponding to the sets of non-zero and zero true coefficient $\beta_i$. Moreover, we take 
\begin{equation*}
\begin{aligned}
      G_{H_i}   &= \{g \in \groupset : g \subseteq H_i \},\\
      G_{H^c_i}   &= \{g:g \subseteq H^c_i \}, \\
      G_{H_{0,i}} &= \{g:|g \cap H_i| > 0; |g \cap H^c_i| > 0\}, 
\end{aligned}
\end{equation*}
the sets of groups in which the indices are respectively all non-zero, all zero and a mix of zero and non-zero in $\beta_i$. To investigate the asymptotic properties of the estimator in (\ref{eq_aogl}), we make the following assumptions:

\begin{Aassumption}
$\plim_{T_i \to \infty} \hat{Q}_{x,i} = Q_{x,i}$, where $\hat{Q}_{x,i} = \frac{1}{T_i}\sum_t I_{i,t} x_{i,t} x_{i,t}^{\top}$ and  $Q_{x,i} = \mathbb{E}[x_{i,t}x_{i,t}^\top|\gamma_i]$ is positive definite.
\label{assum:x_converge}
\end{Aassumption}

\begin{Aassumption}
$\mathbb{E}[\varepsilon_{i,t}|\varepsilon_{i,\underline{t-1}}, \mathcal{F}_{t}] = 0$ with $\varepsilon_{i,\underline{t}} = \{\varepsilon_{i,t}, \varepsilon_{i,t-1}, \ldots \}$ and there exists a positive constant $M$ such that for all $n,T$, $\frac{1}{M} \leq \sigma_{i}^2\leq M,\ i=1,...,n$, with $\sigma_{i}^2 = \mathbb{E}[\varepsilon_{i,t}^2| \gamma_i]$.
\label{assum:mds}
\end{Aassumption}

\begin{Aassumption}
There exists a neighborhood in $\real^d$ around $\beta_i$ such that the decomposition of any vector $b$ in the neighborhood has unique decomposition $\{v_{i,g}^b\}$ minimizing the norm $\|\beta_i\|_{2,1,\groupset}$. In particular, the decomposition $\{v_{i,g}^b\}$, minimizing the norm $\|\beta_i\|_{2,1,\groupset}$ is unique. Further, this decomposition is such that $\{v_{i,0}^b\} = 0$, for all $g \in G_{H_{0,i}}$.
\label{assum:uniqueness}
\end{Aassumption}

 Assumption~\ref{assum:x_converge} is a usual assumption for the standard OLS solution to be consistent (see Assumption B.1 in \citetalias{gagliardini2016time}), while Assumption~\ref{assum:mds}  allows for or a martingale difference sequence for the error terms (see Assumption A.1 in \citetalias{gagliardini2016time}). Assumption~\ref{assum:uniqueness} is discussed in \cite{percival2012theoretical} and addresses the uniqueness of decomposition of $\beta_i$. We now state the main result for the first-pass regression, which corresponds to Theorem 2 derived by \cite{percival2012theoretical} in the fixed design framework. 
\begin{Lemma} (Asymptotic normality of $\hat \beta_i$) \\
Under Assumptions APR.1 to APR.3, SC.1 and SC.2 of \citetalias{gagliardini2016time}, Assumptions~\ref{assum:x_converge} to \ref{assum:uniqueness}, let $\beta_i^{\text{init}}$ be an initial  $\sqrt{T_i}$-consistent estimator and let $\{v^{\text{init}}_{i,g}\} = \mathcal{V}(\beta_i^{\text{init}})$ be any decomposition minimizing the norm $\|\beta_i^{\text{init}}\|_{2,1,\groupset}$. For all $i \in \{1, \ldots, n\}$, let $\delta_g = \nicefrac{1}{\|v^{\text{init}}_{i,g}\|^{\check{\gamma}}}$, for $\check{\gamma}>0$, such that $T_i^{\nicefrac{(\check{\gamma}+1)}{2}} \delta \to \infty$. If $\sqrt{T_i}\delta \to 0$, then, as $T_i \to \infty$, we get the convergence in distribution:
\begin{equation*}
    \sqrt{T_i}\left(\hat{\beta}_{i} - \beta_{i}\right) \Longrightarrow V_i, 
\end{equation*}
where the vector $V_i$ has entries
\begin{equation*}
    \begin{aligned}
        V_{H_i} &\sim N(0,\sigma^2_iQ^{-1}_{H_i,x,i}), \\
        V_{H^c_i} &= 0,
    \end{aligned}
\end{equation*}
where ${Q}_{H_i,x,i}$ is the submatrix of  ${Q}_{x,i}$ with indices in $H_i$.
\label{lemma_asynorm_beta}
\end{Lemma}

For the above result to hold, the vector $\beta_i^{\text{init}}$ needs to be $\sqrt{T_i}$-consistent. More specifically, $\beta_i^{\text{init}}$ is any $a_{T_i}$-consistent estimator where  $a_{T_i} \to \infty$, and $a_{T_i}^{\check{\gamma}} \sqrt{T_i} \delta \to \infty$. Moreover, in the context of the aOGL, the decomposition  $\{v^{\text{init}}_{i,g}\}$ must be unique. Lemma 4 in \cite{percival2012theoretical} shows $\sqrt{T_i}$-consistency of the $\{v^{\text{OLS}}_{i,g}\}$, which is a example of a potential solution for $\{v^{\text{init}}_{i,g}\}$ in the case of fixed covariates. In our framework, with the uniqueness assumption of the decomposition, we can use the ridge regression estimator as  $\{v^{\text{init}}_{i,g}\}$.
The distributional result of Lemma \ref{lemma_asynorm_beta} is key in deriving the asymptotic properties of the second-pass regression discussed in the next section.

To control for short sample size, and potentially numerical instability on the inversion of matrix $\hat{Q}_{x,i}$, we consider the trimming device defined in \citetalias{gagliardini2016time}, such that $\boldsymbol{1}_{i}^{\chi}=\boldsymbol{1}\{ CN(\hat{Q}_{x,i})\leq\chi_{1,T},\tau_{i,T}\leq\chi_{2,T}\} $,
where $ CN(\hat{Q}_{x,i})=\sqrt{\eig_{\max}(\hat{Q}_{x,i})/\eig_{\min}(\hat{Q}_{x,i})}$
is the condition number of the matrix $\hat{Q}_{x,i}$, $\eig_{\min}(\cdot)$ denotes the minimum eigenvalue, and $\tau_{i,T}=T/T_{i}$. The first trimming based on $CN(\hat{Q}_{x,i})\leq\chi_{1,T}$ selects the assets for which the time-series regression is not badly conditioned, while the second trimming based on  $\tau_{i,T}\leq\chi_{2,T}$ keeps only the assets for which samples are not too short. 

\subsection{Second-pass regression}
\label{sec:secondpass}

The second-pass regression aims at computing the cross-sectional estimator of $\nu$. For that purpose, we implement the WLS estimator of \citetalias{gagliardini2016time}, while accounting for the sparse model specification in the first-pass regression for all $i = 1, \ldots, n$. 
For that purpose, we introduce the indicator vector $ \mathbf{1}_{\beta_i} \in \natural^d$, such that $\mathbf{1}_{\beta_{i,l}} = 1$ if $\beta_{i,l} \neq 0$, and $0$ otherwise, for $l = 1, \ldots, d$, that we decompose in the following manner:
$
     \mathbf{1}_{\beta_i} = (\mathbf{1}_{{\beta}_{11,i}}^{\top}, \mathbf{1}_{{\beta}_{12,i}}^{\top}, \mathbf{1}_{{\beta}_{21,i}}^{\top}, \mathbf{1}_{{\beta}_{22,i}}^{\top})^{\top},
$
where $\mathbf{1}_{{\beta}_{11,i}} \in \natural^{d_{11}}$, $\mathbf{1}_{{\beta}_{12,i}} \in \natural^{d_{12}}$, $\mathbf{1}_{{\beta}_{21,i}} \in \natural^{d_{21}}$ and  $\mathbf{1}_{{\beta}_{22,i}} \in \natural^{d_{22}}$. To implement the WLS estimator for the vector $\nu$, we need to account for the different number of regressors selected through the aOGL approach. Hence, in the same spirit as in \cite{chaieb2020factors}, we introduce the following selection matrices that  help us transforming the $x_{i,t}$ into their sparse counterparts.  The matrices $\tilde{D}_i$ and $\tilde{E}_i$ are  the $d_{11} \times d_{11,i}$ and $d_{12} \times d_{12,i}$ such that columns with all zeros have been removed in $\diag[\mathbf{1}_{{\beta}_{11,i}}]$ and $\diag[\mathbf{1}_{{\beta}_{12,i}}]$. Similarly, the matrices $\tilde{B}_i$ and $\tilde{C}_i$ are  the $d_{21,i} \times d_{21}$ and $d_{22,i} \times d_{22}$ matrices such that rows with all zeros have been removed in $\diag[\mathbf{1}_{{\beta}_{21,i}}]$ and $\diag[\mathbf{1}_{{\beta}_{22,i}}]$. Moreover, we define $x_{H_i,i,t}$ as the vector of regressors indexed by $H_i$ after the selection of the first pass.

Based on the selection matrices $\tilde{D}_i,\tilde{E}_i,\tilde{B}_i$, and $\tilde{C}_i$ , we rewrite the parameter restriction in (\ref{eq_condNoArbRest}) such that 
\begin{equation*}
\begin{aligned}
        {\beta}_{1,i} &= \biggl(\tilde{D}_i^{\top} N_{\tilde{p}} \left[\left({\Lambda} - {F}\right)^{\top}\otimes I_{\tilde{p}}\right]\tilde{B}_i^{\top}\tilde{B}_i \vect\left[\breve{B}_i^{\top}\right],\\ 
    &\quad \quad \tilde{E}_i^{\top} W_{\tilde{p},q}\left[\left({\Lambda} - {F}\right)^{\top}\otimes I_q\right] \tilde{C}_i^{\top}\tilde{C}_i\vect\left[C_i^{\top}\right]\biggr)^{\top}, \\   
   {\beta}_{3,i} &= \left(\left[\tilde{D}_i^{\top} N_{\tilde{p}}\left(\breve{B}_i^{\top}\otimes I_{\tilde{p}}\right)\right]^{\top}, \left[ \tilde{E}_i^{\top}W_{\tilde{p},q}\left(C_{i}^{\top}\otimes I_{p}\right)\right]^{\top}\right)^{\top},
\end{aligned}
\end{equation*}
where $N_{\tilde{p}}$ is defined in (\ref{betaone}), yielding the asset pricing restrictions expressed in the newly defined ${\beta}_{1,i}$ and ${\beta}_{3,i}$ as
$
    {\beta}_{1,i} = {\beta}_{3,i} \nu,$ $ \nu=\vect[\Lambda^{\top}-F^{\top}].
    $
We obtain ${\beta}_{3,i}$ from the following identity,
\begin{equation*}
\small
    \begin{aligned}
        \vect[{\beta}_{3,i}^{\top}]&=J_{a,i}{\beta}_{2,i},\\
        J_{a,i}&=\left(\begin{array}{cc}
                        J_{11,i} & 0\\
                            0 & J_{22,i}
                        \end{array}\right),\\
        J_{11,i}&=W_{d_{11,i},Kp}\left[I_{Kp}\otimes\left(\tilde{D}_i^{\top}N_{\tilde{p}}\right)\right] \left\{I_{K} \otimes \left[\left(W_{p}\otimes I_{p}\right)\left(I_{p}\otimes \vect\left[I_{p}\right]\right)\right]\right\}\tilde{B}_i^{\top},\\
        J_{22,i}&=W_{d_{12,i},Kp}\left[I_{Kp}\otimes\left(\tilde{E}_i^{\top}W_{p,q}\right)\right]\left\{ I_{K} \otimes \left[\left(W_{p,q}\otimes I_{p}\right)\left(I_{p}\otimes \vect\left[I_{q}\right]\right)\right]\right\}\tilde{C}_i^{\top}.
    \end{aligned}
\end{equation*}
We can now implement the following second-pass regression WLS estimator
\begin{equation}
\hat{\nu}=\hat{Q}_{\beta_3}^{-1}\frac{1}{n}\sum_{i}\hat{\beta}_{3,i}^{\top}\hat{w}_{i}\hat{\beta}_{1,i},
\label{nuWLScond}
\end{equation}
 where $\hat{\nu}$ denotes the estimator of $\nu$, $\hat{Q}_{\beta_{3}}=\frac{1}{n}\sum_{i}\hat{\beta}_{3,i}^{\top}\hat{w}_{i}\hat{\beta}_{3,i}$ , and weights are estimates of $w_{i}=\boldsymbol{1}_{i}^{\chi}\left(\diag\left[v_{i}\right]\right)^{-1}$.
Moreover, the $v_{i}$ are the asymptotic variances of the standardized errors $\sqrt{T}(\hat{\beta}_{1,i}-\hat{\beta}_{3,i}\nu)$
in the cross-sectional regression for large $T$ such that $ v_{i}=\tau_{i}C_{\nu,1,i}^{\top}Q_{H_i,{x},i}^{-1}S_{ii}Q_{H_i,{x},i}^{-1}C_{\nu,1,i}$, where $ S_{ii}=\plim_{T\to\infty}\frac{1}{T}\sum_{t}\sigma_{i}^2{x}_{H_i,i,t}{x}_{H_i,i,t}^{\top}$  and $C_{\nu,1,i}$ $=$ $(E_{1,i}^{\top}-(I_{d_{1,i}}\otimes\nu^{\top})J_{a,i}E_{2,i}^{\top})^{\top}$, $E_{1,i}=(I_{d_{1,i}},0_{d_{1,i}\times d_{2,i}})^{\top}$, $E_{2,i}=(0_{d_{2,i}\times d_{1,i}},I_{d_{2,i}})^{\top}$.
We use the estimates $ \hat{v}_{i}=\tau_{i,T}C_{\hat{\nu}_{1}}^{\top}\hat{Q}_{H_i,{x},i}^{-1}\hat{S}_{ii}\hat{Q}_{H_i,{x},i}^{-1}C_{\hat{\nu}_{1}}$, where $ \hat{S}_{ii}=\frac{1}{T_{i}}\sum_{t}I_{i,t}\hat{\varepsilon}_{i,t}^{2}{x}_{H_i,i,t}{x}_{H_i,i,t}^{\top}$,
$\hat{\varepsilon}_{i,t}=R_{i,t}-\hat{\beta}_{i}^{\top}{x}_{H_i,i,t}$
together with  $ C_{\hat{\nu},1,i}=(E_{1,i}^{\top}-\left(I_{d_{1,i}}\otimes\hat{\nu}_{1,i}^{\top}\right)J_{a,i}E_{2,i}^{\top})^{\top}$.
To estimate $C_{\nu,1,i}$, we use the OLS estimator given by $ \hat{\nu}_{1,i}=(\sum_{i}\boldsymbol{1}_{i}^{\chi}\hat{\beta}_{3,i}^{\top}\hat{\beta}_{3,i})^{-1}\sum_{i}\boldsymbol{1}_{i}^{\chi}\hat{\beta}_{3,i}^{\top}\hat{\beta}_{1,i}$. We estimates the weights with $\hat{w}_{i}=\boldsymbol{1}_{i}^{\chi}\left(\diag\left[\hat{v}_{i}\right]\right)^{-1}$.

To study the asymptotic properties of the estimator $\hat{\nu}$, we consider the following assumption on the size of the cross-section $n$.
%
%
\begin{Aassumption}
The size of the cross-section is such that $n = \mathcal{O}(T^{\bar{\gamma}})$ for $\bar{\gamma} > 0$.
\label{assum_wrong_model}
\end{Aassumption}

%
Assumption~\ref{assum_wrong_model} puts a bound on the growth of the cross-section such that it does not grow faster that some power of the sample size $T$. In Proposition~\ref{theo_consitency_nu}, we provide the consistency result for the estimator $\hat{\nu}$. 
\begin{Proposition} (Consistency of $\hat{\nu}$) \\
Under Assumptions APR.1 to APR.4, SC.1 and SC.2, B.1 of \citetalias{gagliardini2016time} and Assumptions~\ref{assum_spec_fact_load}, \ref{assum_spec_riks_premia}, \ref{assum:x_converge} to  \ref{assum_wrong_model}, and \ref{Assum_Qxx} to \ref{assum_order_max}, we have that 
$
   \Vert \hat{\nu} - {\nu} \Vert = o_p\left(1\right),
$
when  $n,T \to \infty$.
\label{theo_consitency_nu}
\end{Proposition}
Assumptions~\ref{Assum_Qxx} to \ref{assum_order_max} are discussed in Appendix~\ref{append_regularity}. This asymptotic property of $\hat{\nu}$ is studied under the double asymptotics $n,T \to \infty$ in  \citetalias{gagliardini2016time}. They show consistency of $\hat{\nu}$ under a full representation of $\beta_i$, while we assume a sparse representation of $\beta_i$. Hence, our result differs in that respect.

Let us now recover the sparse structure of the conditional expectation of  the factors under Assumption~\ref{assum_spec_riks_premia}. For that purpose, we consider the aLASSO estimator of \cite{zou2006adaptive} to select and estimate the matrix $F$ of coefficients. We solve the following minimization problem for all factor $f_{k,t},k = 1, \ldots, K$, such that the estimator of the $k$-th row of the matrix $F$ is given by:
\begin{equation}
    (\hat{F}_{0,k}, \hat{F}_{1,k}) = \argmin_{(F_{0,k},F_{1,k})   \in \real^{\tilde{p}}}  \sum_t \left(f_{k,t} -  F_{0,k} - F_{1,k} Z_{t-1} \right)^2 + \delta  \sum^{p}_{j = 1} \hat{w}_j |F_{1,k,j}|, 
    \label{eq:flasso}
\end{equation}
where $\delta$ accounts for the overall amount of shrinkage as in (\ref{eq_aogl}), and $\hat{w_j}$ are data dependent weights. Typically, the weights are defined as $\hat w_j = \nicefrac{1}{|\hat{F}_{1,k,j}^{\text{OLS}}|^{\check{\gamma}}}$ for $\check \gamma >0$, where $\hat{F}_{1,k,j}^{\text{OLS}}$ are the OLS estimates of $F_{1,k,j}$, the true values in the vector parameter $F_{1,k}$. The estimate $\hat{F}$ stacks row-wise the elements of $(\hat{F}_{0,k}, \hat{F}_{1,k})$ obtained from (\ref{eq:flasso}). Under Assumption~\ref{assum_time_ivariant}, no amount of shrinkage is applied to $F_0$ in $F$, to always keep the time-invariant contribution in the model. We get the final estimates of the sparse matrix $\Lambda$ from the relationship $ \vect[\hat{\Lambda}^{\top}]=\hat{\nu}+\vect[\hat{F}^{\top}]$, which yields $\hat{\lambda}_{t}=\hat{\Lambda}Z_{t-1}$. To derive the asymptotic consistency of $\hat{\Lambda}$, we rely on Proposition~\ref{theo_consitency_nu} for the estimator $\hat{\nu}$. Let us consider the following assumption:
\begin{Aassumption} We have
    $\plim_{T \to \infty} \nicefrac{1}{T} \sum_{t=1}^T \tilde{Z}_{t-1}\tilde{Z}_{t-1}^{\top}$ 
    $= \mathbb{E}[\tilde{Z}_{t-1} \tilde{Z}_{t-1}^\top]$, where $\mathbb{E}[\tilde{Z}_{t-1} \tilde{Z}_{t-1}^\top]$ is a positive definite matrix. 
\label{assum_knight}
\end{Aassumption}
Assumptions \ref{assum_knight} is a standard regularity assumption on the design matrix for linear regression model, in order to obtain a unique solution for $(F_{0,k},F_{1,k})$. Under the above Assumption~\ref{assum_knight}, and Proposition~\ref{theo_consitency_nu}, the following proposition gives the consistency result for the estimator $\hat{\Lambda}$.
\begin{Proposition}(Consistency of $\hat{\Lambda}$)\\
Under Assumptions  APR.1 to APR.4, SC.1 and SC.2, B.1 of \citetalias{gagliardini2016time}, Assumptions~\ref{assum_spec_fact_load}, \ref{assum_spec_riks_premia}, \ref{assum:x_converge} to  \ref{assum_knight} and \ref{Assum_Qxx} to \ref{assum_bound_zt}, we have that 
$
   \Vert \hat{\Lambda} - {\Lambda} \Vert = o_p\left(1\right),
$
when  $n,T \to \infty$.
\label{theo_consitency_lambda}
\end{Proposition}
Proof of Proposition~\ref{theo_consitency_lambda} is direct since from the definition of $\hat{\Lambda}$,   $\Vert \vect [\hat{\Lambda}^{\top} - \Lambda^{\top}]\Vert \leq \Vert \hat{\nu}  - \nu \Vert  +  \Vert \vect[\hat{F}^{\top} - F^{\top}]\Vert$. From Proposition \ref{theo_consitency_nu}, we know that $\Vert \hat{\nu}  - \nu \Vert = o_p(1)$. Moreover, the aLASSO estimator in (\ref{eq:flasso}) is a special case of the estimator in (\ref{eq_aogl}), where each group is a singleton. Hence, considering Assumptions~\ref{assum_knight} and \ref{assum_bound_zt} which are the counterpart of Assumptions~\ref{assum:x_converge} and \ref{assum:mds} respectively, we get that the result of Lemma~\ref{lemma_asynorm_beta} applies to (\ref{eq:flasso}). Hence, $\Vert \vect[\hat{F}^{\top} - F^{\top}]\Vert = o_p(1)$. Therefore, we get consistency of $\hat{\lambda}_t$, $\sup_t\Vert \hat{\lambda}_t - \lambda_t \Vert = o_p(1)$, under Assumptions~\ref{assum_knight} and \ref{assum_bound_zt}.

\section{Simulation study}
\label{sec:MC}

In this section, we study how the selection and estimation procedures of Section~\ref{sec_estim} perform in finite samples. This first simulation study aims at investigating the prediction and selection performance of the aOGL method and at comparing it with the aLASSO method  in a very sparse environment (Assumptions~\ref{assum_spec_fact_load} and \ref{assum_spec_riks_premia}). To that purpose, we simulate $500$ replicates from the DGP in (\ref{eq_dgp}) for a (randomly drawn) single asset $i$ with sample size $T_i = 500$. We split that full sample in a training subsample and a testing subsample of $450$ and $50$ observations. The testing set is used for out-of-sample prediction performance assessment, where we compare the realized excess returns $R_{i,t}$ with their predictions $\hat{R}_{i,t} = \hat b_{i,t}^{\top} \hat \lambda_t$ under the model estimated on the training set. Errors in (\ref{eq_dgp}) are i.i.d.\ such that $\varepsilon_{i,t}\sim\mathcal{N}(0,\sigma^2)$, where $\sigma = 0.09$. We match the model specification described in our empirical study (Section~\ref{sec_data}) for the common instruments $Z_{t-1} \in \real^6$ and stock-specific instruments $Z_{i,t-1}\in \real^{13}$. For the factors, we use the Fama-French five-factor model \citep{fama2015five} described in the next section, namely we condition w.r.t.\ the values $f_t$ observed in our empirical study for the five factors.
We also condition w.r.t.\ the observed $Z_{t-1}$ and $Z_{i,t}$ for asset $i$ of our empirical study. We only draw the error terms as in a parametric bootstrap. 


In accordance with sparsity in Assumptions~\ref{assum_spec_fact_load} and \ref{assum_spec_riks_premia} and non-sparse time-invariant contribution in Assumption~\ref{assum_time_ivariant}, we set the matrices $A_i$, $B_i$, and $C_i$ according to their values for asset $i$ in the empirical study, with one non-zero element in $B_i$ and two non-zero element for $C_i$. We keep the vector $A_i$ full.
We set the corresponding $a_{i,t}$ in order to avoid \textit{ex-ante} arbitrage. Since we take very sparse matrices $B_i$ and $C_i$, we can view the simulation study as conservative for selection performance assessment (type of worst-case scenario). The resulting $\beta_i$ has 28 non-zero coefficients (including the 6 coefficients induced by the non-sparse time-invariant contribution) over a total of 219 coefficients.  The matrices $F$ and $\Lambda$ are simply set to zero since they do not concern the aOGL estimator.

%
%

%
%
The selection and prediction performance is measured through the average Root Mean Squared Prediction Error (Av($ \text{RMSPE}_{R} $)), the average Root Mean Squared Error for parameter $\beta_i$ (Av($\text{RMSE}_{{\beta}}$)), the proportion of times the model introduces arbitrage (Arb.\ ($\%$)), the average number of selected true non-zero coefficients (True+), and average number of regressors in the selected model (NbReg). Table \ref{tab_simu1} summarizes the results. The aOGL method makes a better job at predicting out-of-sample with a reduction of 1.7\% w.r.t.\ the aLASSO method. The improvement in the average of RMSE for $\beta_i$ is 109\%.  The standard errors are also much lower (reduction of 9.1\% and 91.0\% for the Av($ \text{RMSPE}_{R}$) and Av($\text{RMSE}_{{\beta}}$)). Contrary to the aLASSO method, for which $98.2\%$ of estimated models exhibit  arbitrage, the aOGL method selects only models without introducing \textit{ex-ante} arbitrage  by construction.  Since we face less than $100\%$ for the aLASSO method, it sometimes shrinks  adequately to zero  the coefficients that should be. The aOGL method  is able to recover in average the 11 true non-zero coefficients (11.26) while the aLASSO method struggles (7.37). The aOGL method is also more parsimonious than the aLASSO method in terms of selected regressors (average of 14.75 versus 16.05).

\begin{table}[!ht]
	\centering
	\begin{tabular}{lcccccc}
	\hline
	\centering Method & Av($ \text{RMSPE}_{R}$) & Av($\text{RMSE}_{{\beta}}$) & Arb.\ ($\%$) & True+  & NbReg\\
	\hline
	aOGL &  $9.60 \cdot 10^{-2}$  & $1.48 \cdot 10^{-3}$ & $0.0$ &  $11.26$ &  $14.75$\\
	& $(4.38 \cdot 10^{-4})$ & $(9.33 \cdot 10^{-6})$ & ( - ) & (0.20)  & (0.31) \\
	aLASSO &  $9.76 \cdot 10^{-2}$  & $3.10 \cdot 10^{-3}$ & $98.2$ & $7.37$ &  $16.05$ \\
	& $(4.82 \cdot 10^{-4})$ & $(1.04 \cdot 10^{-4})$ &  (1.12)  &  (0.10) & (0.61) \\
	\hline
	\end{tabular}
	\caption{Performance of estimation and model selection criteria. The methods include the aOGL and aLASSO. We simulate 500 samples under the true sparse DGP. We report the average Root Mean Squared Prediction Error (Av($ \text{RMSPE}_{R}$)), the average Root Mean Squared Error for parameter $\beta_i$ (Av($\text{RMSE}_{{\beta}}$)), the proportion of times the model does not introduce arbitrage (Arb.\ ($\%$)), the average number of selected true non-zero coefficients (True+), and the average number of regressors in the selected model (NbReg),  with their respective standard errors in parenthesis.}
	\label{tab_simu1}
\end{table}

\begin{figure}
    \centering
    \includegraphics[scale = .6]{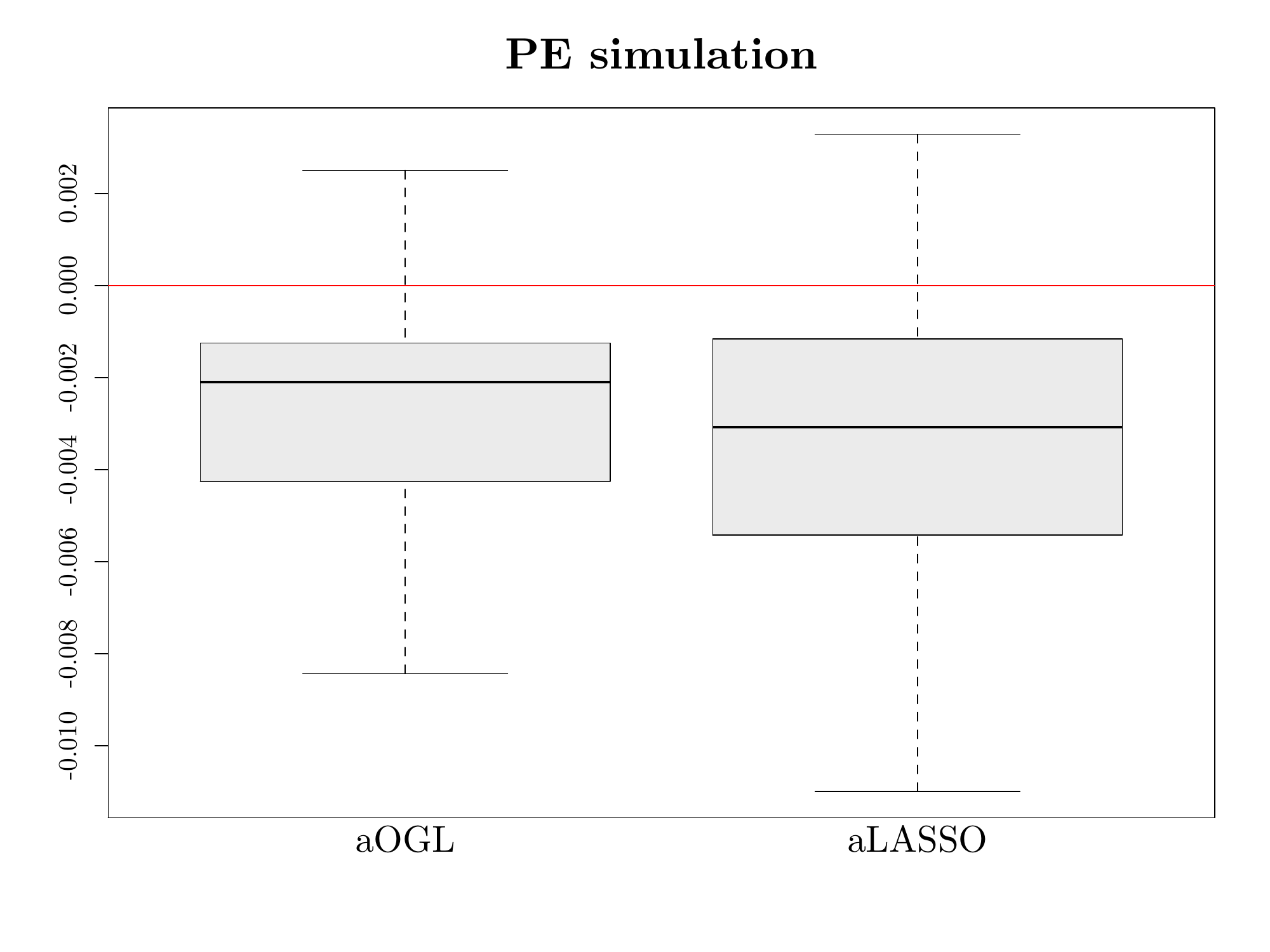}
    \caption{Empirical distribution of out-of-sample Prediction Error (PE) of an equally-weighted portfolio. We compare the aOGL and aLASSO methods.  We simulate excess return paths for 500 assets under sparse DGPs.
    }
    \label{fig_simu}
\end{figure}

 Our second simulation set-up focuses on the out-of-sample prediction performance of the aOGL method in a setting close to our empirical study of Section \ref{sec_perf_out}. We use a training sample to estimate the model and a testing sample to gauge its out-of-sample prediction performance on an equally-weighted portfolio. We consider the same model specification in terms of $f_t$, $Z_{t-1}$ and $Z_{i,t-1}$ as in the first study and implement the following procedure. We sample randomly a subset of $n = 500$ assets from Section~\ref{sec:empirical_result} (training sample), while keeping the same proportion of time-invariant models as in Table~\ref{tab_arbitrage}. From each asset $i$ in this subset, we simulate $T_i$ observations from $R_{i,t} = a_{i,t} + b_{i,t}^{\top} f_t + \varepsilon_{i,t}$ with the coefficients $a_{i,t}$ and $b_{i,t}$ chosen as their aOGL corresponding values for stock $i$. The $500 \times 1$ error vector $\varepsilon_{t}$ at date $t$ is Gaussian with mean zero and block-diagonal correlation matrix with 10 blocks of equal size $50$, where, within each block matrix, the correlation between $\varepsilon_{k,t}$ and $\varepsilon_{l,t}$ is set to $\text{corr}(\varepsilon_{k,t},\varepsilon_{l,t}) =  0.25^{|k-l|}$, $k,l= 1,..., 50, l \neq k$. The variance of each error $\varepsilon_{i,t}$ is set equal to 0.05. From those 500 simulated paths, we implement the aOGL estimation procedure of Section~\ref{sec_estim}, and compare it with the same procedure, but using the aLASSO estimator instead of the aOGL estimator to select the covariates in (\ref{eq_LinearFactorRegression}).
\begin{table}[!ht]
\centering
\begin{tabular}{@{}lccc@{}}
\toprule
  Methods & Av(RMSPE) & Av(MAPE) \\
 \cmidrule(r){1-1} \cmidrule(lr){2-2} \cmidrule(lr){3-3} \cmidrule(lr){4-4}  
 aOGL &  $6.39 \cdot 10^{-2} $& $ 4.19 \cdot 10^{-2}$    \\
 &  $(2.18 \cdot 10^{-3} ) $& $ (7.01 \cdot 10^{-4})$    \\
 aLASSO & $9.89 \cdot 10^{-2}$ & $ 4.87 \cdot 10^{-2}$    \\
  & $(6.90 \cdot 10^{-3} ) $& $ (4.30 \cdot 10^{-3})$    \\
\bottomrule
\end{tabular}
 \caption{Out-of-sample prediction performance of an equally-weighted portfolio. We compare the aOGL  and  aLASSO methods. We simulate excess return paths for 500 assets under  sparse DGPs. We report the average of Root Mean Squared Prediction Error (Av(RMSPE)), and the average of Mean Absolute Prediction Error (Av(MAPE)) of an equally-weighted portfolio with their respective standard error{\color{red}s} in parenthesis.}
\label{tab_simu2}
\end{table}
To evaluate the out-of-sample prediction performance, we simulate one new cross-sectional sample (testing sample) from the time-varying factor model for the 500 assets and each date $t$ and compute the prediction $\hat{R}_{i,t} = \hat b_{i,t}^{\top} \hat \lambda_t$ for the 500 stocks and each date $t$ based on the estimator computed before through the aOGL and aLASSO methods. We finally compute the out-of-sample 
Prediction Error (PE) for an  equally-weighted  portfolio through the difference between the new simulated $\frac{1}{500} \sum_i R_{i,t}$ and  its predicted value $\frac{1}{500} \sum_i \hat R_{i,t}$.  We compute the Root Mean Squared Prediction Error (RMSPE), and the Mean Absolute Prediction Error (MAPE) over the vector gathering the PE at each out-of-sample date. We repeat this procedure 100 times to get an average and to compute a standard error. They are reported in Table~\ref{tab_simu2}. We can see that the aOGL method is much better at out-of-sample predicting excess  returns of an equally-weighted portfolio both in terms of average of  MAPE (reduction by 14\%) but also in terms of variability as measured by the standard errors (reduction by 84\%). The empirical distribution of the prediction errors is given in Figure~\ref{fig_simu}. We can see that the aOGL method is centered closer to zero and with a lower dispersion when compared to the aLASSO method.
Those second simulation results again point in favor of our advocated estimation method.

\section{Empirical results}
\label{sec:empirical_result}
This section investigates the predictive capacity of the aOGL estimator and compares it with the aLASSO estimator. We also consider a pure time-invariant model, and a (hybrid) model with constant $\nu$ and time-varying risk premia. We use the aLASSO estimator to gauge the added value of incorporating the no-arbitrage restrictions in the penalisation approach and the time-invariant models to  gauge the added value of allowing for full time-variation.   The latter comparison checks that, when it comes to return prediction, the complicated model does not necessarily outperform because of potential overfitting.

\subsection{Data description}
\label{sec_data}

We extract the stock returns from the CRSP database for US common stocks listed on the NYSE, AMEX, and NASDAQ, and remove stocks with prices below 5 USD. We exclude financial firms (Standard Industrial Classification Codes between 6000 and 6999). The firm characteristics come from COMPUSTAT. The sample begins in July 1963 and ends in December 2019. It gives us $T = 678$ monthly observations. We proxy the risk-free rate with the 1-month T-bill rate. 

From \cite{freyberger2020dissecting}, we consider the following $q=13$ firm level characteristics $Z_{i,t-1}$: change in share outstanding ($\Delta$ shrout), log change in the split adjusted shares outstanding ($\Delta$ so), growth rate in total assets (Inv), size (LME), last month volume over shares outstanding (lturnover), adjusted profit margin (PM), momentum and intermediate momentum ($r_{12,2}$ and $r_{12,7}$), short-term reversal ($r_{2,1}$), closeness to 52-week high (Rel\_to\_high), the ratio of market value of equity plus long-term debt minus total assets to Cash and Short-Term Investments (ROC), standard unexplained volume (SUV), and  total volume (Tot\_vol). We refer to \cite{freyberger2020dissecting} for a detailed description of those characteristics. We only retain stocks for which all 13 characteristics are non-missing. It produces a sample of $n = 6874$. For each $Z_{i,t-1}$, we follow \cite{freyberger2020dissecting} and compute the cross-sectional rank at each time $t-1$ for all observations \citep[see also][]{chaieb2020factors}. For the common instruments $Z_{t-1}$, we consider the $p=6$ following variables: dividend yield (dp), net equity expansion (ntis), inflation (infl), stock variance (svar), default spread (def\_spread), and the term-spread (term\_spread).  For each $Z_{t-1}$, we center and standardize all observations.

We consider the two following sets of factors $f_t$. The first set is the four-factor model of \cite{carhart1997persistence}, such that  $f_t = (f_{m,t}, f_{hml,t}, f_{smb,t}, f_{mom,t})^{\top}$, where $f_{m,t}$ is the month $t$ market excess return over the risk free rate, $f_{hml,t}$, $f_{smb,t}$, $f_{mom,t}$ are respectively the month $t$ returns on zero investment factor-mimicking portfolio for size, book-to-market, and momentum. Our second set of factors considers the profitability factor $f_{rmw,t}$ and the investment factor $f_{cma,t}$ as in the five-factor model of \cite{fama2015five}, such that 
$f_t = (f_{m,t}, f_{hml,t}, f_{smb,t}, f_{rmw,t},f_{cma,t})^{\top}$. Our choice for a parsimonious specification in the factor space is justified by our goal of studying the selection of common and idiosyncratic instruments $Z_{t-1}$ and $Z_{i,t-1}$ that have impacts on the dynamics of the $a_{i,t}$, $b_{i,t}$, and $\lambda_t$.  \cite{gagliardini2019diagnostic} and \cite{gagliardini2019estimation} also report evidence that those factors with time-varying loadings are rich enough to achieve a weak cross-sectional dependence in the error terms, namely there are no remaining omitted factors in the error terms.

\subsection{In-sample prediction performance and selection results}
\label{sec_perf_in}

In this section, we investigate the selection results from the first-pass penalized regression. We compare the fit of the penalized two-pass procedure with aOGL described in Section~\ref{sec_estim} to the aLASSO estimator, where we select the $x_{i,t}$ and estimate their coefficients in the first-pass regression with the aLASSO estimator of \cite{zou2006adaptive} and fit the WLS estimator for the $\nu$ described in Section~\ref{sec_model}. We  compute the estimator $\hat{F}$ as in (\ref{eq:flasso}). The horse race starts from the same set of initial data described in the previous section, and the comparison is thus made on the same initial full information. From the characteristics and common instruments outlined in Section~\ref{sec_data}, under the  Carhart four-factor model, we have  $d=5$ for the time-invariant model  and $d=199$ for the time-varying model. Regarding the five-factor model of \cite{fama2015five}, we have $d=6$ and $d=219$ for the unconditional and conditional specifications. The number of possible models under the aLASSO method is $2^{194}$ ($2^{213}$) with $K=4$ ($K=5$), while the number of possible models under the aOGL method is $2^{97}$ ($2^{116}$), which gives the ratio $2^{-97}$, a much lower value than the upper bound $1/8$ in (\ref{upper_bound}).

We choose the regularisation parameter in a data dependent way to minimize the Akaike Information Criterion (AIC) for both aOGL and aLASSO estimator. As advocated in \cite{greene2008econometric}, we use $\chi_{1,T} = 15$, and require at least 5 years of data such that $\chi_{2,T} = 678/60$.
Because of the trimming, we do not keep the same set of stocks for each method and each model. Indeed, due to the different models induced by the first pass for each stock $i$, the trimming device $\boldsymbol{1}\{CN(\hat{Q}_{\check{x},i})\leq\chi_{1,T}\},$ yields a different set of stocks for each method. Since we do not wish to introduce multicolinearity in the second-pass regression, we choose to stick with different sets for each method. For the aOGL estimator, the aLASSO estimator, and the time-invariant estimator, we end up with 4412, 2225, 4879 for the four-factor model, and 4441, 2097, 4879 for the five-factor model. We can observe that the trimming device for the aLASSO method is more binding.  As seen in the simulation results in Section~\ref{sec:MC} and
in Table~\ref{tab_arbitrage}, the aLASSO method tends to include more variables, and, as a consequence, increase its associated condition number. Table \ref{tab_arbitrage} reports the percentage (TI ($\%$)) of estimated models shrunk towards the time-invariant models. For those estimates,  we only select the single group corresponding to Restriction~\ref{res:ti} related to Assumption~\ref{assum_time_ivariant}.
Around two thirds of the stocks require dynamics in their factor loadings. This new empirical result based on a penalization approach illustrates the relevance of allowing for potential time-variation in modelling excess returns of individual stocks with factor models.
Table \ref{tab_arbitrage} also reports the percentage (Arb. ($\%$)) of estimated models with time-varying loadings and presenting arbitrage, namely selecting covariates violating the no-arbitrage restrictions. For that computation, both the time-invariant estimates and aOGL estimates avoid \textit{ex-ante} arbitrage by construction. In line with our Monte Carlo results, the aLASSO procedure ends up with all the time-varying models estimated with arbitrage for both factor specifications.  We conclude that the aOGL estimation achieves parsimony while avoiding arbitrage in time-varying factor models.
\begin{table}[hbt!]

\centering
\begin{tabular}{@{}lcccccc@{}}
\toprule
  & \multicolumn{3}{c}{Carhart four-factor} & \multicolumn{3}{c}{Fama-French five-factor} \\
\cmidrule(lr){2-4}  \cmidrule(l){5-7}
  Methods & TI ($\%$)& Arb. ($\%$) & Av NbReg & TI ($\%$) & Arb. ($\%$) & Av NbReg \\
 \cmidrule(r){1-1} \cmidrule(lr){2-2} \cmidrule(lr){3-3} \cmidrule(lr){4-4}  \cmidrule(lr){5-5} \cmidrule(lr){6-6} \cmidrule(l){7-7} 
 aOGL & $38$ & $0$ & $13.24$ & $35$  & $0$ &  $14.15$ \\
 aLASSO & $46$  & $100$ &  $33.45$ & $31$  & $100$ & $37.20$ \\
 time-invariant &  $100$ & $0$  & $5$ & $100$ & $0$ & $6$\\

\bottomrule
\end{tabular}
\caption{Percentage (TI $(\%)$) of estimated models shrunk towards the time-invariant specification, percentage (Arb. ($\%$)) of estimated time-varying models presenting arbitrage and average number of regressors selected (Av NbReg) with the Carhart four-factor and Fama-French five-factor models for the aOGL, aLASSO, and time-invariant methods. The sample of US equity excess returns begins in July 1963 and ends in December 2019.}
\label{tab_arbitrage}
\end{table}

\begin{table}
\centering
\begin{tabular}{@{}lcccccc@{}}
\toprule
\multicolumn{7}{c}{Carhart four-factor} \\
\cmidrule(){1-7}  
 $T_i$ & $\leq$ 6y & 6y - 10y & 20y - 30y & 30y - 40y & 40y - 50y & $\geq$ 50y \\
 \cmidrule(r){1-1} \cmidrule(lr){2-2} \cmidrule(lr){3-3} \cmidrule(lr){4-4}  \cmidrule(lr){5-5} \cmidrule(lr){6-6} \cmidrule(ll){7-7} 
 Nber of stocks &  480 & 1535  & 1542  & 921 & 393 & 406 \\
 Av. \# of sel. var. & 10.89 & 10.80 & 11.45 & 13.21 & 14.32 & 28.81 \\
 TI $(\%)$ & 42.08 & 44.36 & 38.59 & 32.25 & 27.23 & 23.65 \\
dp $(\%)$ & 13.12 & 12.83 & 15.89 & 25.52 & 41.22 & 59.36\\
ntis $(\%)$ & 16.04 & 17.79 & 27.89 & 40.61 & 46.06 & 40.89 \\
infl $(\%)$ & 27.71 & 26.58 & 25.88 & 32.25 & 39.44 & 40.39 \\
svar $(\%)$ & 21.04 & 18.76 & 18.94 & 15.20 &  6.87 & 30.79\\
def\_spread $(\%)$ & 22.08 & 23.71 & 31.97 & 34.96 & 42.24 & 49.51\\
term\_spread $(\%)$ &  34.58 & 31.73 & 29.83 & 35.94 & 31.04 & 27.59\\
\cmidrule(){1-7}  
$\Delta$ shrout $(\%)$ & 0.21  &  0.26  & 0.39  & 0.54 &  0.25  & 10.34 \\
$\Delta$ so $(\%)$& 0.00  & 0.13  & 0.32 &  0.98 &  0.51 &  10.34\\
Inv $(\%)$ & 0.00 & 0.39 & 0.32 & 0.54 & 0.51 & 7.64 \\
LME $(\%)$ & 0.63 & 0.33 & 0.26 & 0.87 & 0.25 & 6.65 \\
lturnover $(\%)$ & 0.83 & 0.52 & 0.39 & 0.98 & 0.25 & 7.64\\
PM $(\%)$ & 0.21 & 0.20 & 0.26 & 0.54 & 0.51 & 6.65\\
$r_{12,2}$ $(\%)$ & 0.63  & 0.59 &  0.26 &  0.87 &  0.51 & 10.59\\
$r_{12,7}$ $(\%)$ & 0.21 &  0.20  & 0.26 &  0.11 &  0.00 & 10.10 \\
$r_{2,1}$ $(\%)$ & 0.00 & 0.13 & 0.19 & 0.76 & 0.51 & 7.39 \\
Rel\_to\_high $(\%)$ &  0.00 & 0.39 & 0.32 & 0.22 & 0.25 & 7.39 \\
ROC $(\%)$ & 0.63 & 0.33 & 0.19 & 0.76 & 0.25 & 6.16 \\
SUV $(\%)$ & 0.63 & 0.26 & 0.13 & 0.65 & 0.00 & 7.14 \\
Tot\_vol $(\%)$ & 0.00 & 0.13 & 0.26 & 0.54 & 0.51 & 6.40 \\
 \bottomrule
\end{tabular}
\caption{Selection results sorted by sample size ($T_i$) for the Carhart four-factor specification. We first report  the number of stocks (Nber of stocks), the average number of selected variables  (Av.\ \# of sel.\ var.) and the percentage of estimated  models  shrunk  towards  the  time-invariant  specification (TI (\%)) per sample size range.  Then we give the percentage w.r.t.\ the total number of stocks  of each of the 6 variables in $Z_{t-1}$ and the 13 variables in $Z_{i,t-1}$ per stock excess return sample size. The sample of US equity excess returns begins in July 1963 and ends in December 2019.  }
\label{tab_ti_c4f}
\end{table}

\begin{table}
\centering
\begin{tabular}{@{}lcccccc@{}}
\toprule
\multicolumn{7}{c}{Fama-French five-factor} \\
\cmidrule(){1-7}  
 $T_i$ & $\leq$ 6y & 6y - 10y & 20y - 30y & 30y - 40y & 40y - 50y & $\geq$ 50y \\
 \cmidrule(r){1-1} \cmidrule(lr){2-2} \cmidrule(lr){3-3} \cmidrule(lr){4-4}  \cmidrule(lr){5-5} \cmidrule(lr){6-6} \cmidrule(ll){7-7} 
 Nber of stocks &  480 & 1535  & 1542  & 921 & 393 & 406 \\
 mean \# of sel. var. & 12.71 & 12.46 & 12.88 & 14.14 & 15.09 & 24.76 \\
 TI $(\%)$ & 39.79 & 40.59 & 36.12 & 32.46 & 30.79 & 19.70 \\
 dp $(\%)$ & 15.00 & 13.29 & 15.30 & 22.80 & 32.32 & 49.26\\
ntis $(\%)$ & 18.12 & 20.91 & 29.57 & 39.31 & 45.04 & 51.97 \\
infl $(\%)$ & 31.46 & 30.81 & 26.07 & 33.33 & 36.64 & 48.28 \\
svar $(\%)$ & 24.79 & 21.50 & 18.74 & 15.20 &  6.87 & 24.63\\
def\_spread $(\%)$ & 25.62 & 25.86 & 32.49 & 36.26 & 40.20 & 46.80\\
term\_spread $(\%)$ &  35.83 & 34.40 & 32.04 & 36.48 & 34.61 & 40.64\\
\cmidrule(){1-7} 
$\Delta$ shrout $(\%)$& 0.42 & 0.33 & 0.65 & 0.98 & 0.51 & 4.93 \\
$\Delta$ so $(\%)$& 0.42 & 0.20 & 0.13 & 0.76 & 0.51 & 5.67\\
Inv $(\%)$ & 0.42 & 0.33 & 0.52 & 0.43 & 0.25 & 4.68 \\
LME $(\%)$ & 0.42 & 0.39 & 0.32 & 0.98 & 0.51 & 4.19 \\
lturnover $(\%)$ & 0.63 & 0.39 & 0.58 & 1.09 & 0.51 & 0.23\\
PM $(\%)$ & 0.00 & 0.00 & 0.65 & 0.65 & 0.51 & 3.69\\
$r_{12,2}$ $(\%)$ & 1.04 & 0.46 & 0.84 & 0.76 & 0.25 & 5.17\\
$r_{12,7}$ $(\%)$ & 0.21 & 0.20 & 0.52 & 0.65 & 0.25 & 4.43 \\
$r_{2,1}$ $(\%)$ & 0.42 & 0.20 & 0.06 & 0.43 & 0.25 & 4.19 \\
Rel\_to\_high $(\%)$ & 0.42  & 0.33 & 0.52 & 0.33 & 0.25 & 4.68\\
ROC $(\%)$ & 0.42 & 0.33 & 0.32 & 0.98 & 0.51 & 3.94 \\
SUV $(\%)$ & 0.42 & 0.26 & 0.39 & 0.98 & 0.25 & 5.17  \\
Tot\_vol $(\%)$ &0.00 & 0.00 & 0.58 & 0.54 & 0.51 & 3.69 \\
\bottomrule
\end{tabular}
\caption{Selection results sorted by sample size ($T_i$) for the Fama-French five-factor specification. We first report  the number of stocks (Nber of stocks), the average number of selected variables  (Av.\ \# of sel.\ var.) and the percentage of estimated  models  shrunk  towards  the  time-invariant  specification (TI (\%)) per sample size range.  Then, we give the percentage w.r.t.\ the total number of stocks  of each of the 6 variables in $Z_{t-1}$ and the 13 variables in $Z_{i,t-1}$ per stock excess return sample size. The sample of US equity excess returns begins in July 1963 and ends in December 2019. }
\label{tab_ti_ff5}
\end{table}

\begin{table}
\centering
\begin{tabular}{@{}lcccc@{}}
\toprule
 \multicolumn{5}{c}{Carhart four-factor} \\
\cmidrule(){1-5}  
  & $f_{m}$ & $f_{hml}$ & $f_{smb}$ & $f_{mom}$ \\
 \cmidrule(lr){2-2} \cmidrule(lr){3-3} \cmidrule(lr){4-4}  \cmidrule(l){5-5}
dp $(\%)$ & 18.44 & 14.66 & 15.41 & 14.59 \\
ntis $(\%)$ & 23.19 & 19.28 & 17.59 & 22.12 \\
infl $(\%)$ & 20.68 & 20.59 & 17.27 & 24.82  \\
svar $(\%)$ & 12.67 & 10.23 & 9.22 & 12.48 \\
def\_spread $(\%)$ & 24.43 & 20.78 & 19.93 & 24.82 \\
term\_spread $(\%)$ & 23.52 & 20.94 & 20.52 & 25.70\\
\cmidrule(){1-5} 
$\Delta$ shrout $(\%)$& 0.88  & 0.75 & 0.68 & 0.78  \\
$\Delta$ so $(\%)$& 1.07 &  1.14 & 0.94 & 1.11 \\
Inv $(\%)$ & 0.75 & 0.62  & 0.65 & 0.65  \\
LME $(\%)$ & 0.62 & 0.42  & 0.46 & 0.39 \\
lturnover $(\%)$ & 1.11 & 0.85 & 0.88 & 0.85 \\
PM $(\%)$ & 0.59 & 1.10  & 0.62 & 0.55 \\
$r_{12,2}$ $(\%)$ & 1.17  & 0.85 & 0.85 & 0.85 \\
$r_{12,7}$ $(\%)$ & 1.01  & 0.88 & 0.85 & 2.73 \\
$r_{2,1}$ $(\%)$ & 0.98 & 0.62 & 0.59 & 0.98\\
Rel\_to\_high $(\%)$ & 0.75 & 0.85 & 1.51 & 1.37 \\
ROC $(\%)$ & 0.88 &  0.82 & 0.75 & 0.91  \\
SUV $(\%)$ & 0.94 & 0.78 & 0.75 & 0.88 \\
Tot\_vol $(\%)$ & 0.81 & 0.72 & 0.65 & 0.72 \\
\bottomrule
\end{tabular}
\caption{Selection results for the Carhart four-factor specification. For stocks exhibiting time variation in their factor loadings,  we report the percentage  of each of the 6 variables in $Z_{t-1}$ and the 13 variables in $Z_{i,t-1}$ selected per factor. The sample of US equity excess returns begins in July 1963 and ends in December 2019. }
\label{tab_sel_c4f}
\end{table}

\begin{table}
\centering
\begin{tabular}{@{}lcccc@{}}
\toprule
 \multicolumn{5}{c}{Carhart four-factor} \\
\cmidrule(){1-5}  
  & $f_{m}$ & $f_{hml}$ & $f_{smb}$ & $f_{mom}$ \\
 \cmidrule(lr){2-2} \cmidrule(lr){3-3} \cmidrule(lr){4-4}  \cmidrule(l){5-5}
dp & & $\checkmark$ & $\checkmark$ & $\checkmark$ \\
ntis & $\checkmark$ & $\checkmark$ & $\checkmark$ & $\checkmark$ \\
infl & $\checkmark$ & $\checkmark$ &  &  \\
svar & $\checkmark$ & $\checkmark$ & $\checkmark$ & $\checkmark$\\
def\_spread & $\checkmark$ & $\checkmark$ & $\checkmark$ & $\checkmark$ \\
term\_spread & $\checkmark$ & $\checkmark$ &  & $\checkmark$\\
\bottomrule
\end{tabular}
\caption{Selection results for the the drivers of $\mathbb{E}[f_{t}\vert\mathcal{F}_{t-1}]$ for the  Carhart four-factor specification.  A check denotes
inclusion of a covariate in $Z_t$. The sample begins in July 1963 and ends in December 2019. }
\label{tab_sel_lambda_c4f}
\end{table}

\begin{table}
\centering
\begin{tabular}{@{}lccccc@{}}
\toprule
 \multicolumn{6}{c}{Fama-French five-factor} \\
\cmidrule(){1-6}  
  & $f_{m}$ & $f_{hml}$ & $f_{smb}$ & $f_{rmw}$ & $f_{cma}$ \\
 \cmidrule(lr){2-2} \cmidrule(lr){3-3} \cmidrule(lr){4-4}  \cmidrule(lr){5-5} \cmidrule(l){6-6}
dp $(\%)$ & 16.45 & 12.49 & 13.63 & 8.48 & 8.95 \\
ntis $(\%)$ & 23.43 & 17.87 & 18.34 & 13.14 & 13.09 \\
infl $(\%)$ & 22.71 & 18.41 & 17.84 & 15.08 & 14.20  \\
svar $(\%)$ & 13.73 & 7.56 & 9.65 & 5.21 & 7.05 \\
def\_spread $(\%)$ &  24.95 & 19.51 & 20.34 & 12.83 & 15.15\\
term\_spread $(\%)$ & 26.88 & 19.51 & 21.47 & 14.90 & 15.18\\
\cmidrule(){1-6} 
$\Delta$ shrout $(\%)$& 0.66  & 0.51 & 0.44 & 0.40 & 0.32 \\
$\Delta$ so $(\%)$& 0.73  &  0.51 & 0.60 & 0.40 &  0.47\\
Inv $(\%)$ & 0.32 & 0.44  & 0.35 & 0.23 & 0.32 \\
LME $(\%)$ & 0.32 & 0.16  & 0.22 & 0.05 & 0.19 \\
lturnover $(\%)$ & 0.44 & 0.35 & 0.33 & 0.25 & 0.28\\
PM $(\%)$ & 0.47 & 0.37  & 0.33 & 0.44 & 0.32\\
$r_{12,2}$ $(\%)$ & 0.89  & 0.54 & 0.51 & 0.48 & 0.38\\
$r_{12,7}$ $(\%)$ & 0.63  & 0.51 & 0.54 & 0.44 & 0.47 \\
$r_{2,1}$ $(\%)$ & 0.47 & 0.35 & 0.38 & 0.51 & 0.19 \\
Rel\_to\_high $(\%)$ &  0.60 & 0.57 & 0.57 & 0.66 & 0.41 \\
ROC $(\%)$ & 0.73 &  0.40 & 0.66 & 0.37 & 0.19 \\
SUV $(\%)$ & 0.85 & 0.48 & 0.47 & 0.55 & 0.41 \\
Tot\_vol $(\%)$ & 0.51 & 0.22 & 0.40 & 0.33 & 0.51 \\
\bottomrule
\end{tabular}
\caption{Selection results for the Fama-French five-factor specification. For stocks exhibiting time variation in their factor loadings,  we report the percentage  of each  of the 6 variables in $Z_{t-1}$ and the 13 variables in $Z_{i,t-1}$ selected per factor. The sample of US equity excess returns begins in July 1963 and ends in December 2019. }
\label{tab_sel_ff5}
\end{table}

\begin{table}
\centering
\begin{tabular}{@{}lccccc@{}}
\toprule
 \multicolumn{6}{c}{Fama-French five-factor} \\
\cmidrule(){1-6}  
  & $f_{m}$ & $f_{hml}$ & $f_{smb}$ & $f_{rmw}$ & $f_{cma}$ \\
 \cmidrule(lr){2-2} \cmidrule(lr){3-3} \cmidrule(lr){4-4}  \cmidrule(lr){5-5} \cmidrule(l){6-6}
dp &  & $\checkmark$ & $\checkmark$ & $\checkmark$ & $\checkmark$ \\
ntis &$\checkmark$  & $\checkmark$ & $\checkmark$ & $\checkmark$ & $\checkmark$ \\
infl & $\checkmark$ & $\checkmark$ &  &  &   \\
svar & $\checkmark$ & $\checkmark$ & $\checkmark$ &  &  \\
def\_spread & $\checkmark$ & $\checkmark$ & $\checkmark$ & $\checkmark$ & $\checkmark$\\
term\_spread & $\checkmark$ & $\checkmark$ &  &  & \\
\bottomrule
\end{tabular}
\caption{Selection results for the the drivers of $\mathbb{E}[f_{t}\vert\mathcal{F}_{t-1}]$ for the  Fama-French five-factor specification. A check denotes
inclusion of a covariate in $Z_t$. The sample begins in July 1963 and ends in December 2019. }
\label{tab_lambdat_sel_ff5}
\end{table}

In the three first lines of Tables \ref{tab_ti_c4f} and \ref{tab_ti_ff5}, we  investigate the type of stock excess returns that exhibit time-variation in the their factor loadings. For both factor specifications, the longer the sample size, the more ``action'' is needed for the dynamics of the factor loadings. Indeed, the aOGL method selects a time-invariant model for more than 50\% of the stocks excess returns exhibiting historical data smaller than 10 years, and this for  both factor specifications. On the contrary, 80\% of the models with the longest sample size ($\geq$ 50 years) need time-variation in their factor loadings.

In the next lines of Tables \ref{tab_ti_c4f} and \ref{tab_ti_ff5}, we further look at the selected variables among the 6 in $Z_{t-1}$ and the 13 in $Z_{i,t-1}$. Across all sample sizes $T_i$, the percentages of selected variables in $Z_{t-1}$ are much higher than the percentages of selected variables in $Z_{i,t-1}$. For smaller time spans,  some characteristics are never selected. Therefore, it seems that the common instruments $Z_t$ are key drivers of the time variation of the factor loadings.  It is particularly true for the range $\geq$ 50y. It shows the need of including common instruments that pick up the influence of the business cycles on the factor loading dynamics in larger time spans. While there are no common instruments that are never selected, and this across all time spans, the proportions far below 100\% demonstrate the need to select instruments in a data-driven way. 

In Tables \ref{tab_sel_c4f} and \ref{tab_sel_ff5}, we report the percentage that the 6 variables in $Z_{t-1}$ (scaled factors) and the 13 variables in $Z_{i,t-1}$ are selected through the aOGL method for both factor specifications. The percentages of selected common instruments are similar for the factors $f_{m}$,  $f_{hml}$, and  $f_{smb}$ shared between  the two models. With the Carhart four-factor specification, we need more variables in $Z_{i,t-1}$ to describe the dynamics of the factor loadings in comparison with the Fama-French five-factor specification.  In line with \cite{chaieb2020factors}, the characteristics are not necessarily paired more often with their corresponding factors. The size characteristic LME is more often paired with the market factor $f_{m}$ than with the size factor $f_{smb}$. On the contrary, the momentum characteristics $r_{12,7}$ and $r_{2,1}$  are  often associated with its corresponding factor $f_{mom}$. Finally, Tables \ref{tab_sel_lambda_c4f} and \ref{tab_lambdat_sel_ff5} show that the conditional expectations of the factors $f_{rmw}$ and $f_{cma}$ in the Fama-French five-factor specification need less covariates than for the other factors. The variables ntis and def\_spread are selected for all factors.

Let us now investigate  in-sample predictability performance. As, in \cite{chaieb2020factors},  we decompose the conditional expected return of asset $i$ for month $t$ for both time-varying factor specifications, as:
\begin{equation}
        \mathbb{E}\left[R_{i,t} \vert\mathcal{F}_{t-1}\right]=a_{i,t} -  b_{i,t}^{\top}\nu_t + b_{i,t}^{\top} \lambda_t = a_{i,t} + b_{i,t}^{\top}\mathbb{E}[f_{t}\vert\mathcal{F}_{t-1}].
        \label{eq_er_1}
\end{equation}
For such time-varying specifications, the contribution of the pricing errors $a_{i,t} -  b_{i,t}^{\top}\nu_t$ is often small, revealing that the no-arbitrage restrictions are met for a vast majority of dates. When they are not, \cite{chaieb2020factors} show that incorporating pricing errors, instead of only relying on $b_{i,t}^{\top} \lambda_t$ in (\ref{eq_er_1}),  helps to predict future equity excess returns. Similarly, for the time-invariant models, we decompose the unconditional expected return as:
\begin{equation}
        \mathbb{E}\left[R_{i,t}\right]=a_i - b_i^{\top}\nu + b_i^{\top}\lambda = a_i + b_i^{\top} \mathbb{E}[f_{t}].
        \label{eq_er_2}
\end{equation}
For such time-invariant specifications, the contribution of the pricing errors $a_{i} -  b_{i}^{\top}\nu$ is often large.  We also consider the case of constant $\nu$ and time-varying risk premia $\lambda_t$ ($\lambda_t \& \nu$), for which we decompose the conditional expected return as 
\begin{equation}
        \mathbb{E}\left[R_{i,t} \vert\mathcal{F}_{t-1}\right]=a_i - b_i^{\top}\nu + b_i^{\top}\lambda_t = a_i + b_i^{\top} \mathbb{E}[f_{t}\vert\mathcal{F}_{t-1}].
        \label{eq_er_3}
\end{equation}
 In such a hybrid model \citep{avramov2004stock}, the time-variation in $E[R_{i,t}\vert {\cal F}_{t-1}]$ only comes from the time-variation in $E[f_{t}\vert {\cal F}_{t-1}]$ since $\nu$ is constant because of the no-arbitrage restrictions with constant $b_i$ and $a_i$. 

\begin{figure}
\centering
 \includegraphics[width=1\linewidth]{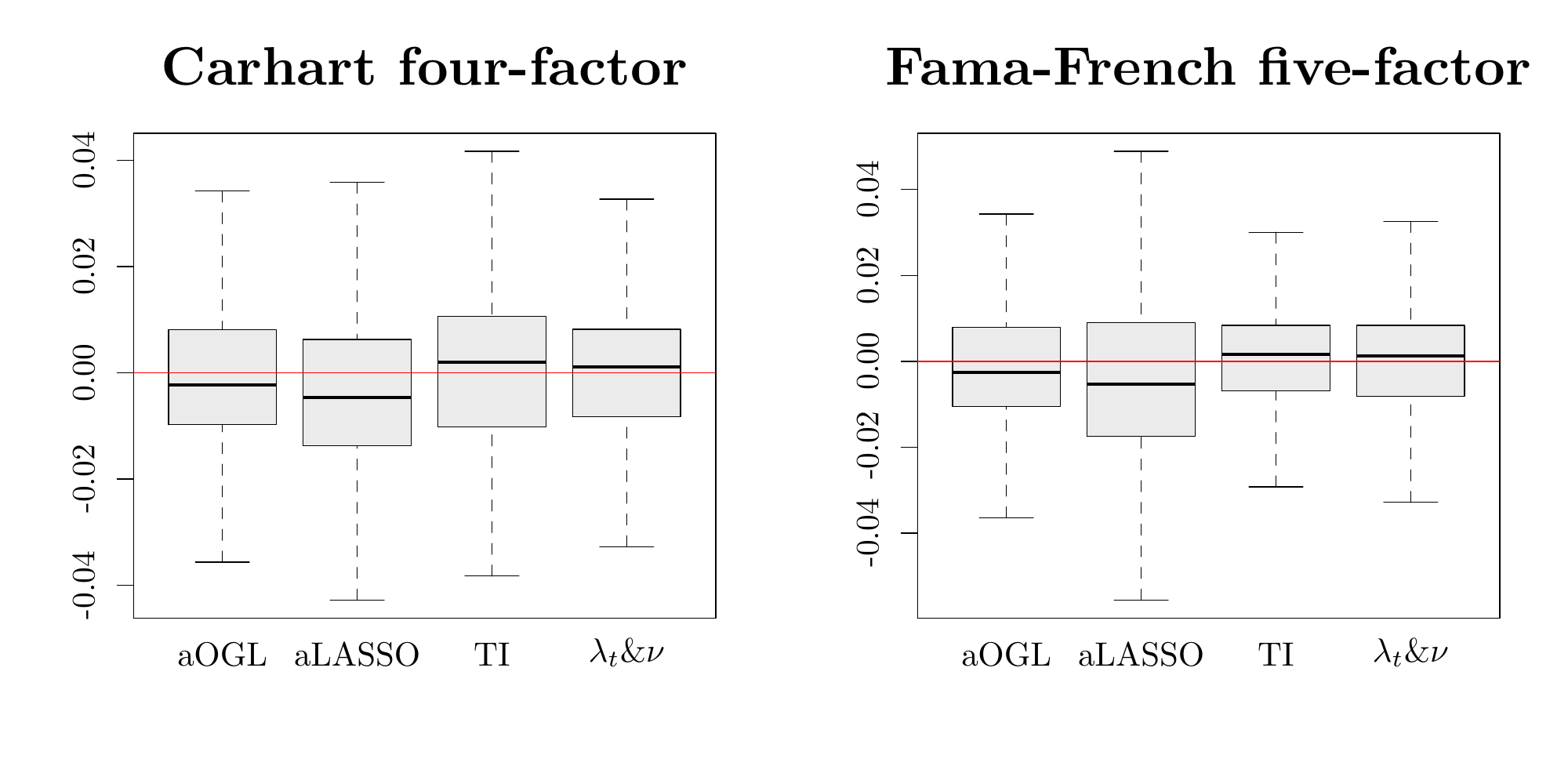}
 \caption{Empirical distribution of in-sample Prediction Error (PE) of an equally-weighted portfolio. We compare the aOGL, aLASSO, time-invariant (TI) and $\lambda_t \& \nu$ methods. The left panel corresponds to the Carhart four-factor model. The right panel corresponds to the Fama-French five-factor model. The sample of US equity excess returns begins in July 1963 and ends in December 2019.}{}
 \label{fig_boxplot_in}
\end{figure}

\begin{figure}
\centering
\begin{subfigure}[b]{\textwidth}
   \includegraphics[width=1\linewidth]{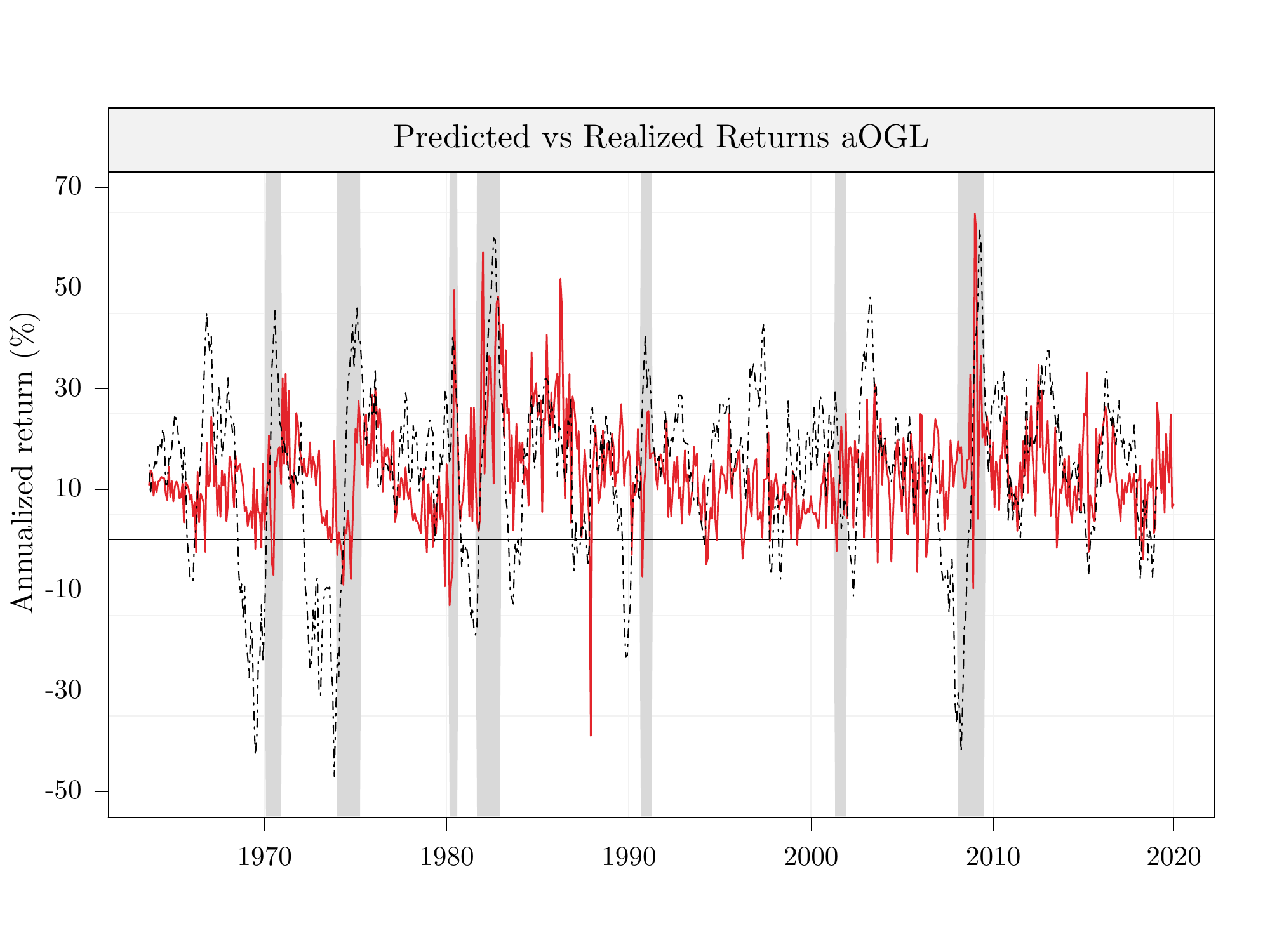}
   \vspace*{-2cm}
\end{subfigure}

\begin{subfigure}[b]{1\textwidth}
   \includegraphics[width=1\linewidth]{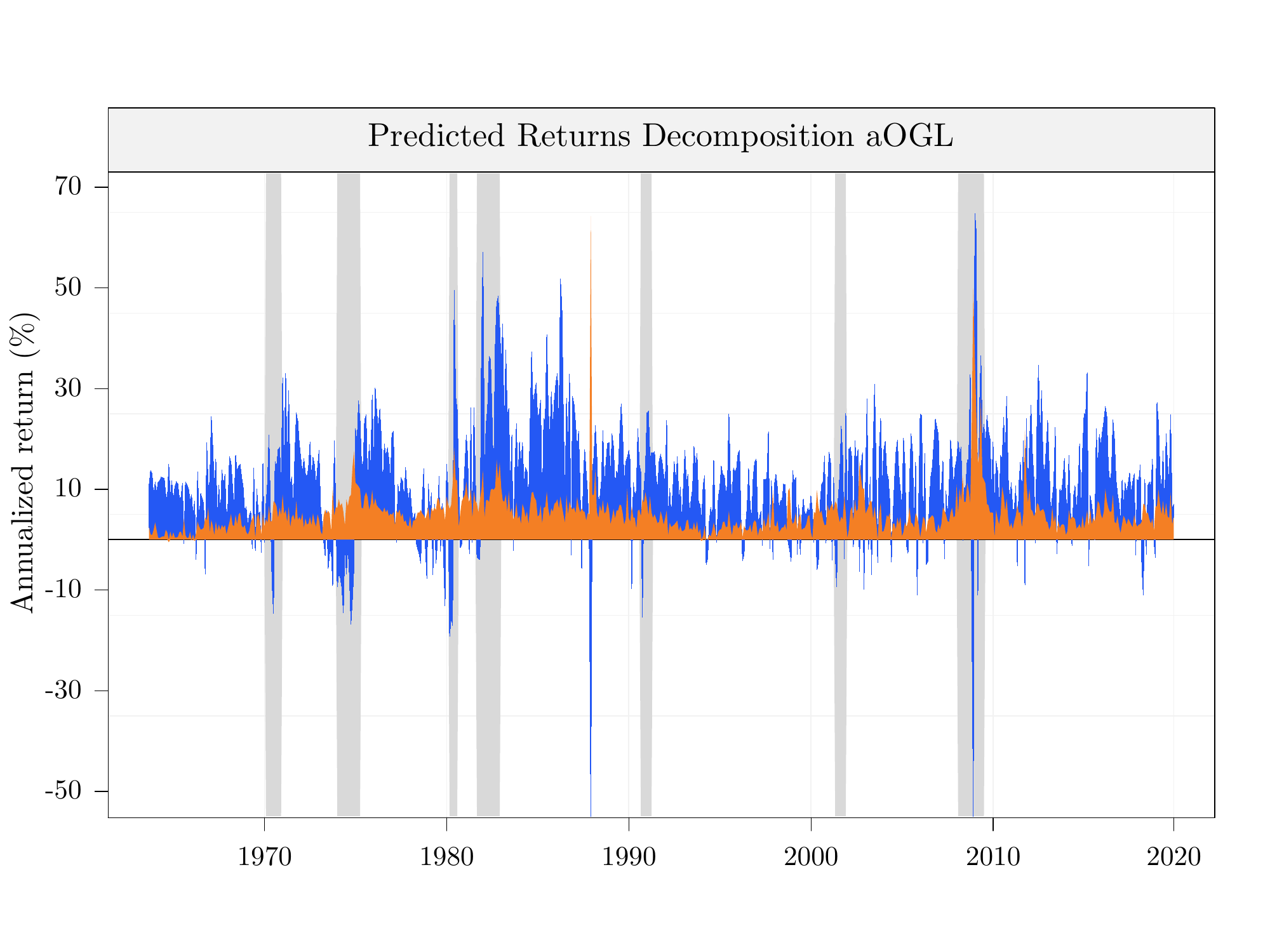}
   \vspace*{-2cm}
\end{subfigure}
\vspace*{0.5cm}
   
 \caption{Predicted excess returns, realized excess returns, and prediction decomposition for the Carhart four-factor model and an equally-weighted portfolio with the aOGL method. In the upper panel, the  predicted excess return path corresponds to the red plain line. The  realized excess returns correspond to the black dashed line. In the lower panel, the orange shaded area corresponds to  estimates of $a_{i,t}$. The blue shaded area corresponds to estimates of $b_{i,t}^{\top}\mathbb{E}[f_{t}\vert\mathcal{F}_{t-1}]$. The gray shaded areas correspond to   the recession periods determined by the National Bureau of Economic Research (NBER). The sample of US equity excess returns begins in July 1963 and ends in December 2019.}{}
 \label{fig_pred_ogl_c4f}
\end{figure}

\begin{figure}
\centering
\begin{subfigure}[h]{\textwidth}
   \includegraphics[width=1\linewidth]{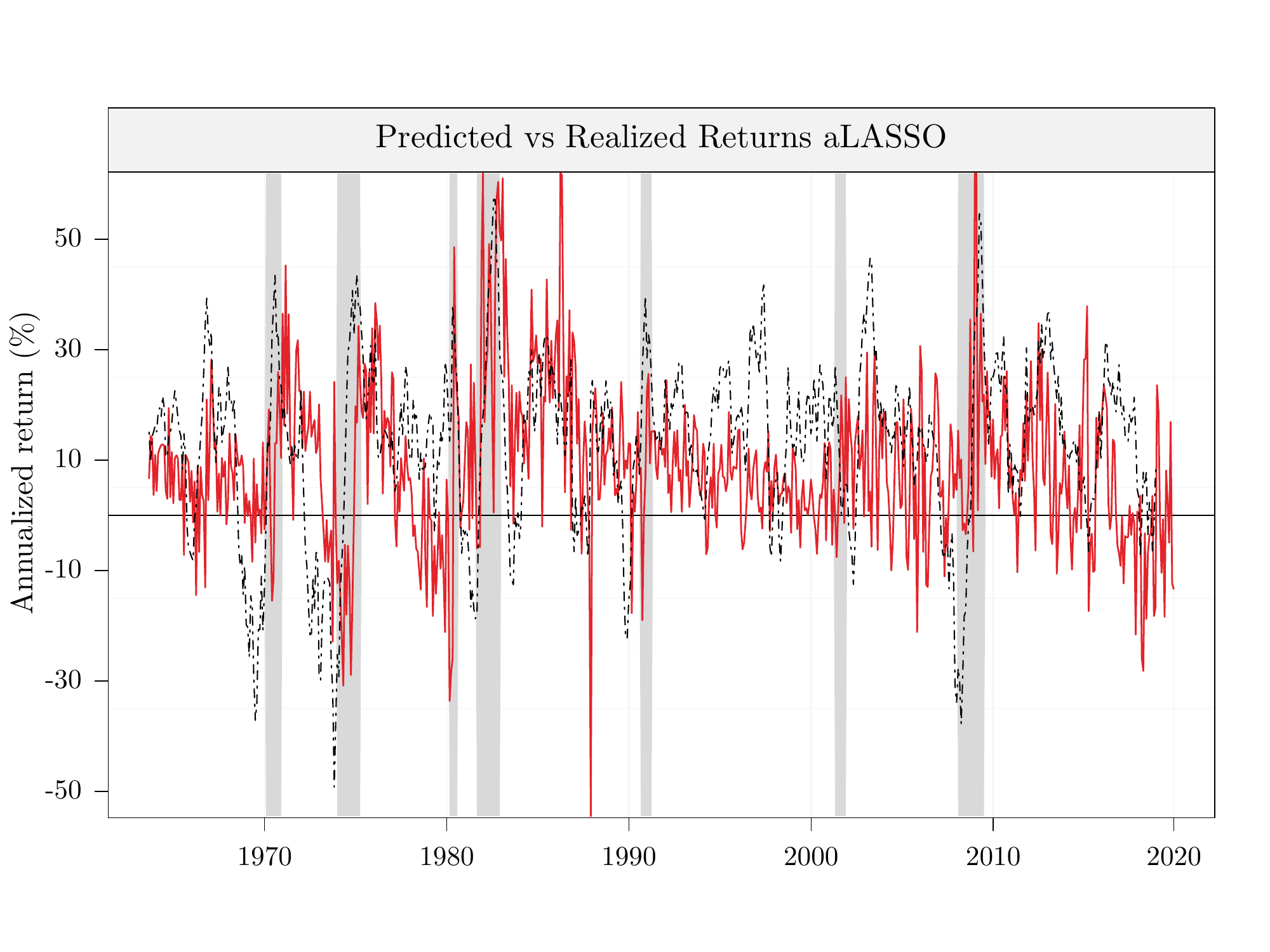}
   \vspace*{-2cm}
\end{subfigure}

\begin{subfigure}[h]{1\textwidth}
   \includegraphics[width=1\linewidth]{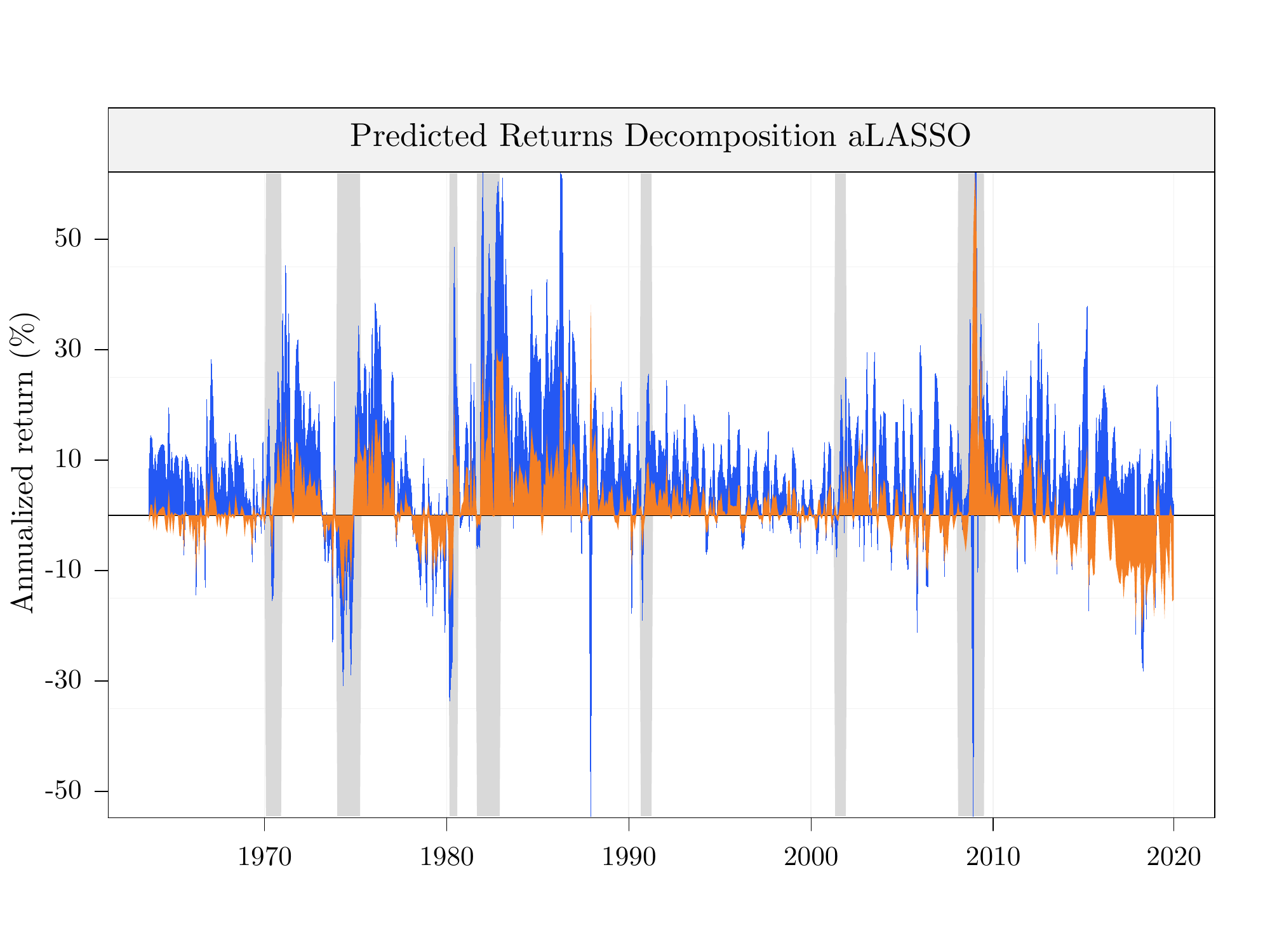}
   \vspace*{-2cm}
\end{subfigure}
\vspace*{0.5cm}
   
 \caption{Predicted excess returns, realized excess returns, and prediction decomposition for the Carhart four-factor model and an equally-weighted portfolio with the aLASSO method. In the upper panel, the  predicted excess return path corresponds to the red plain line. The  realized excess returns correspond to the black dashed line. In the lower panel, the orange shaded area corresponds to   estimates of $a_{i,t}$. The blue shaded area corresponds to estimates of $b_{i,t}^{\top}\mathbb{E}[f_{t}\vert\mathcal{F}_{t-1}]$. The gray shaded areas correspond to   the recession periods determined by the National Bureau of Economic Research (NBER). The sample of US equity excess returns begins in July 1963 and ends in December 2019.}{}
 \label{fig_pred_lasso_c4f}
\end{figure}

\begin{figure}
\centering
\begin{subfigure}[b]{\textwidth}
   \includegraphics[width=1\linewidth]{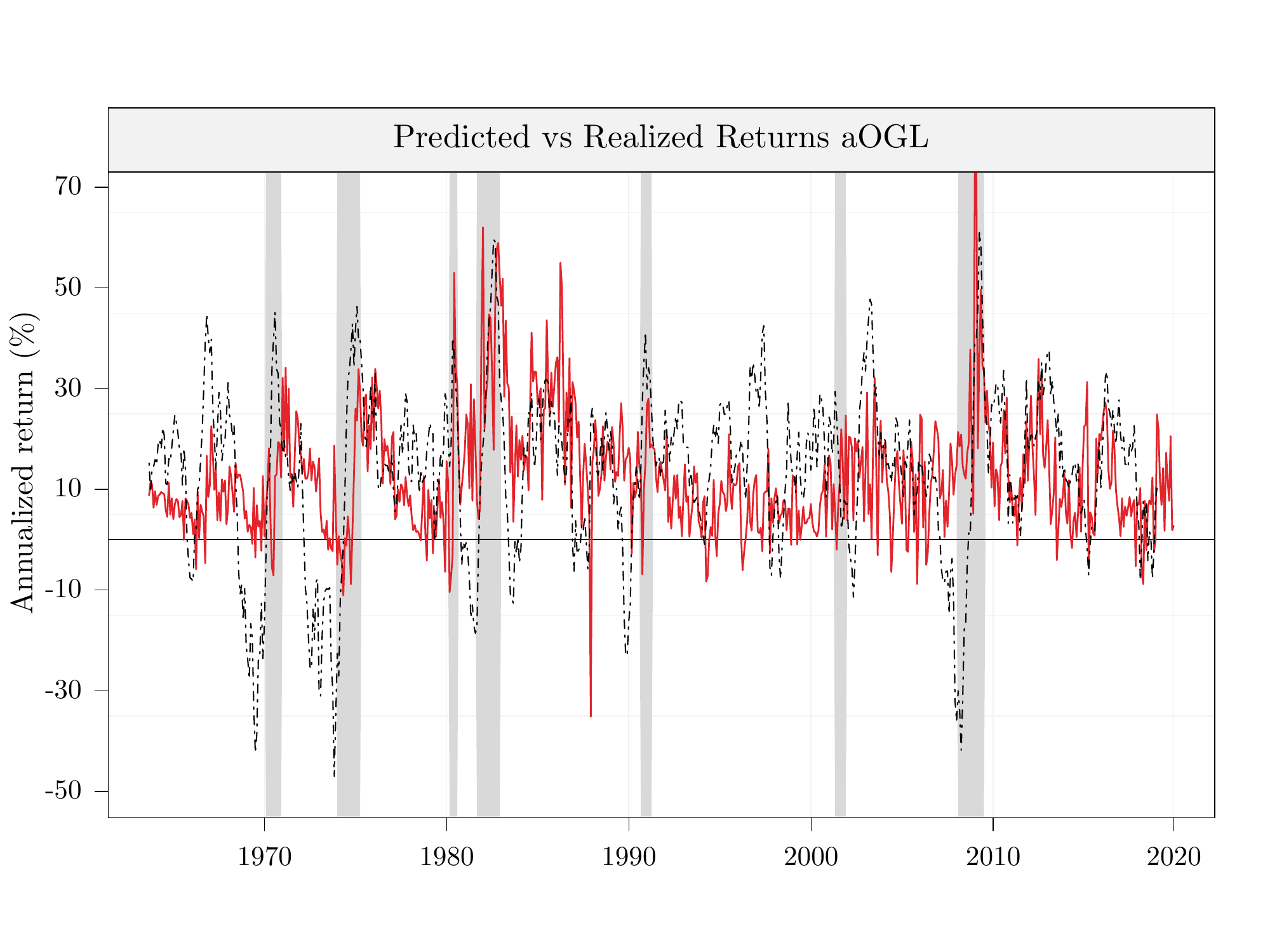}
   \vspace*{-2cm}
\end{subfigure}

\begin{subfigure}[b]{1\textwidth}
   \includegraphics[width=1\linewidth]{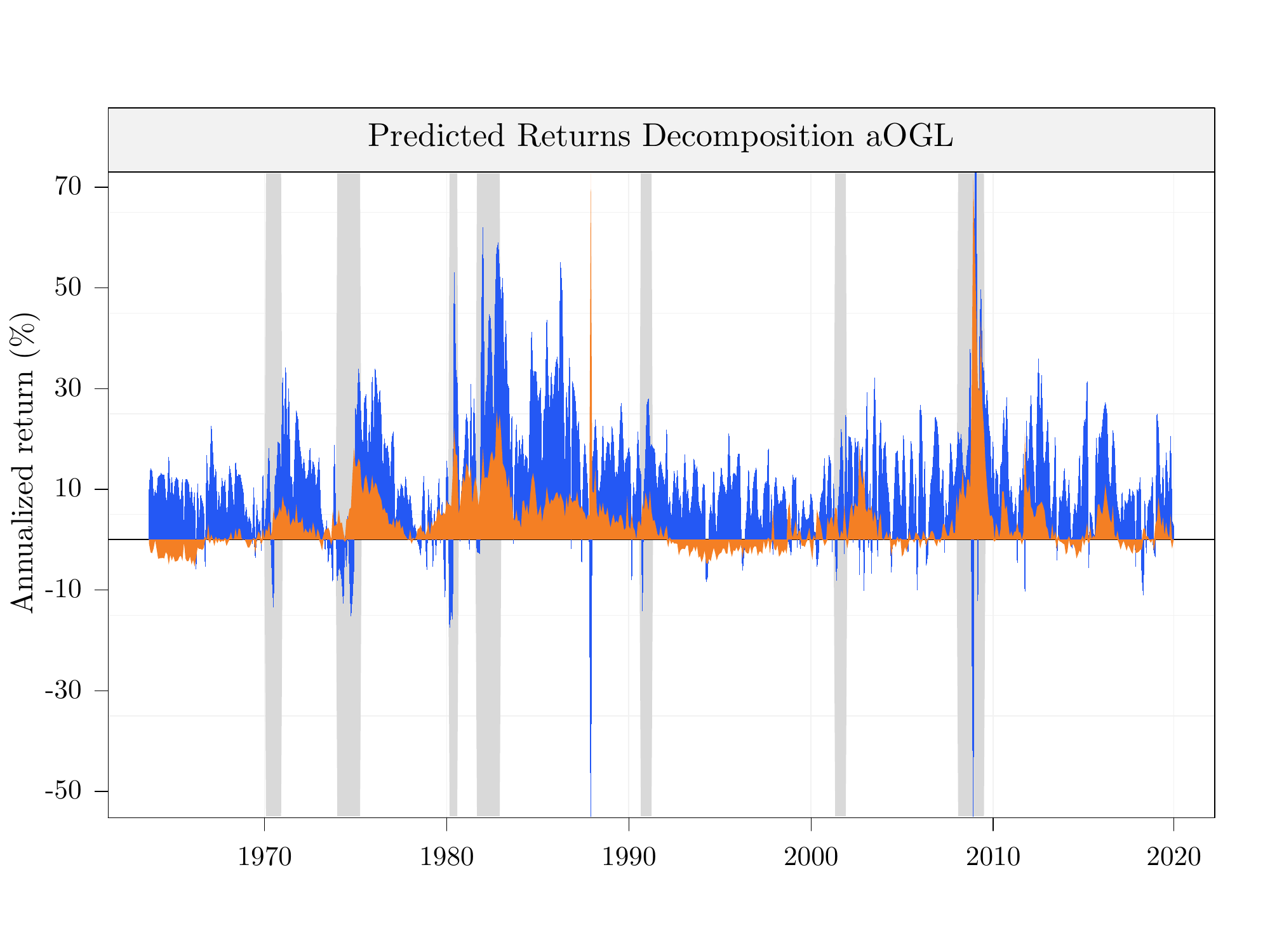}
   \vspace*{-2cm}
\end{subfigure}
\vspace*{0.5cm}
   
 \caption{Predicted excess returns, realized excess returns, and prediction decomposition for the Fama-French five-factor model and an equally-weighted portfolio with the aOGL method. In the upper panel, the  predicted excess return path corresponds to the red plain line. The  realized excess returns correspond to the black dashed line. In the lower panel, the orange shaded area corresponds to  estimates of $a_{i,t}$. The blue shaded area corresponds to estimates of $b_{i,t}^{\top}\mathbb{E}[f_{t}\vert\mathcal{F}_{t-1}]$. The gray shaded areas correspond to   the recession periods determined by the National Bureau of Economic Research (NBER). The sample of US equity excess returns begins in July 1963 and ends in December 2019.}{}
 \label{fig_pred_ogl_ff5}
\end{figure}

\begin{figure}
\centering
\begin{subfigure}[h]{\textwidth}
   \includegraphics[width=1\linewidth]{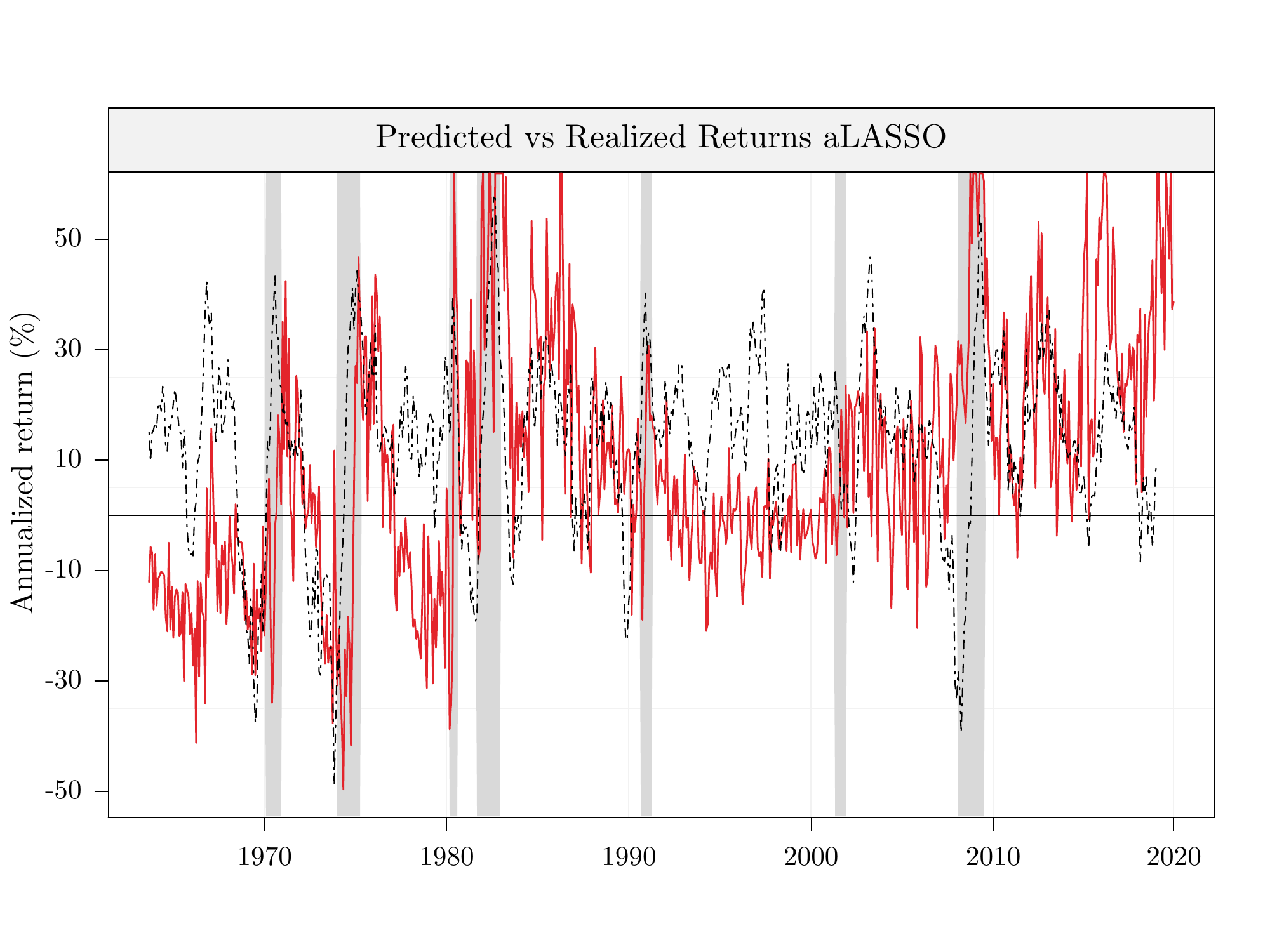}
   \vspace*{-2cm}
\end{subfigure}

\begin{subfigure}[h]{1\textwidth}
   \includegraphics[width=1\linewidth]{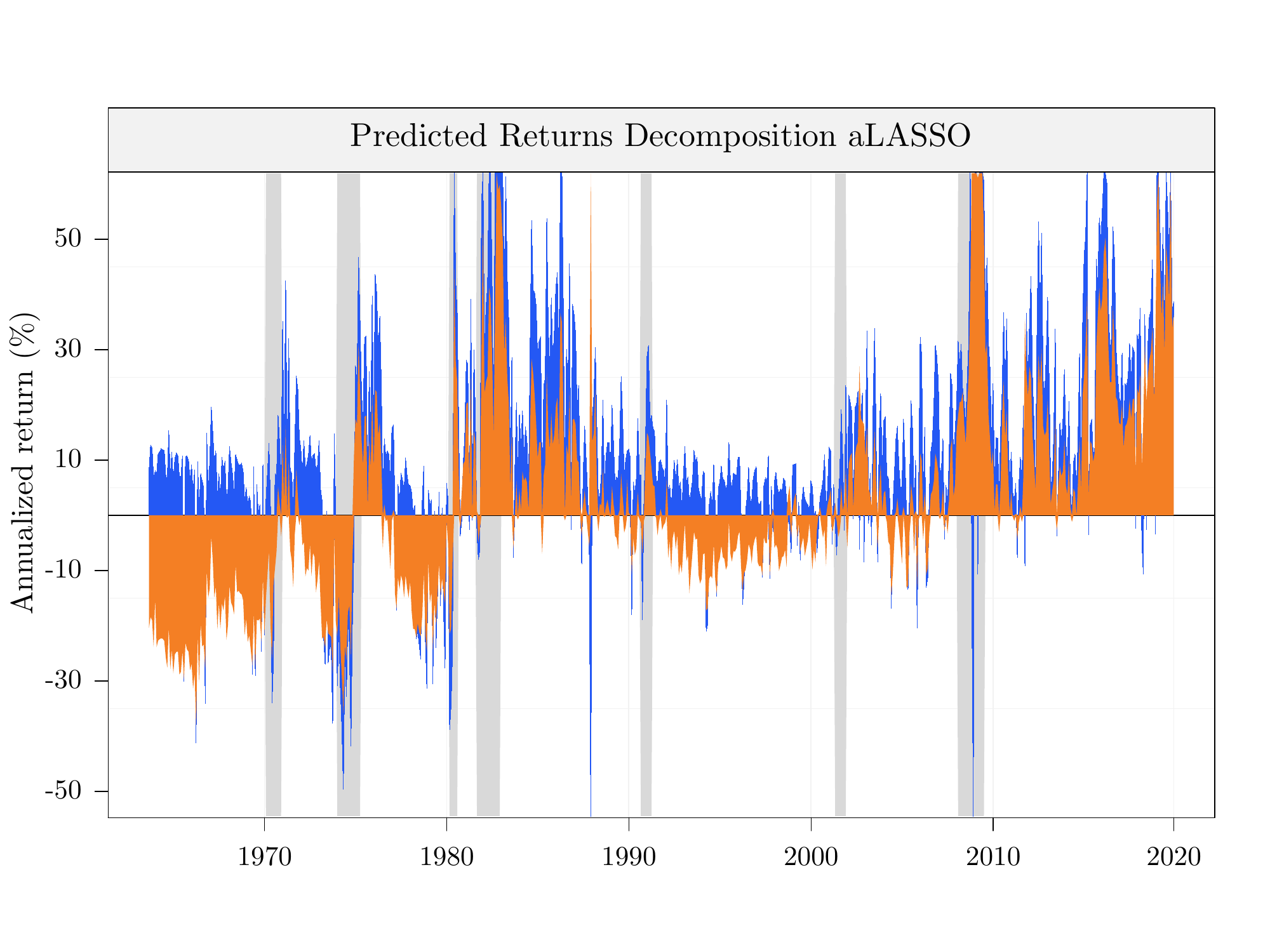}
   \vspace*{-2cm}
\end{subfigure}
\vspace*{0.5cm}
   
 \caption{Predicted excess returns, realized excess returns, and prediction decomposition for the Fama-French five-factor model and an equally-weighted portfolio with the aLASSO method. In the upper panel, the  predicted excess return path corresponds to the red plain line. The  realized excess returns correspond to the black dashed line. In the lower panel, the orange shaded area corresponds to   estimates of $a_{i,t}$. The blue shaded area corresponds to estimates of $b_{i,t}^{\top}\mathbb{E}[f_{t}\vert\mathcal{F}_{t-1}]$. The gray shaded areas correspond to   the recession periods determined by the National Bureau of Economic Research (NBER). The sample of US equity excess returns begins in July 1963 and ends in December 2019.}{}
 \label{fig_pred_lasso_ff5}
\end{figure}

To compare the prediction performance of the four estimation approaches, we compute the RMSPE of an equally-weighted portfolio for the Carhart four-factor model and Fama-French five-factor model. Equal weighting corresponds to cross-sectional averaging. \cite{chaieb2020factors} also uses this weighting scheme. For that portfolio, we compute the PE by comparing the prediction made at time $t$ by each model ((\ref{eq_er_1}) and  (\ref{eq_er_2})) to the forward 12-months realized excess returns, namely the average of the realized excess returns over the next 12 months. Table \ref{tab_rmspe_equal} reports the RMSPE, as well as the Av$(\vert \text{PE}\vert)$  and  Std$(\vert\text{PE}\vert)$ for the Carhart four-factor model and Fama-French five-factor model specifications. The aOGL method performs better than its  natural competitor,  the aLASSO, even for that very diversified stable portfolio, where we expect differences in prediction performance to be attenuated.  It is comparable in terms of the RMSPE to the $\lambda_t \& \nu$ method, with a lower Std$(\vert\text{PE}\vert)$.
Figure~\ref{fig_boxplot_in} displays the corresponding box-plots of the PE computed at each month for each method. The box-plots for the aOGL method in Figure~\ref{fig_boxplot_in} are narrower than for the aLASSO method, and comparable for the two other methods. Those predictability improvements against the aLASSO approach provide further evidence in support for the aOGL approach advocated for the first-pass regression, so that we can incorporate model parameter restrictions to get  models compatible \textit{ex-ante}  with the no-arbitrage restrictions.
To further investigate time-varying predictability, Figures~\ref{fig_pred_ogl_c4f} to \ref{fig_pred_lasso_ff5}  show the forward 12-months realized excess returns for the equally-weighted portfolio and compare them with the predicted excess returns computed from (\ref{eq_er_1}) and (\ref{eq_er_2}) for the two methods with penalisation, respectively for the Carhart four-factor and Fama-French five-factor specifications.
In both Figures~\ref{fig_pred_ogl_c4f} and \ref{fig_pred_ogl_ff5}, the aOGL predicted excess return paths (red plain line) overall reconcile well with the realized excess returns (black dashed line). On the contrary, the aLASSO method in Figures~\ref{fig_pred_lasso_c4f} and \ref{fig_pred_lasso_ff5} does not reconcile well the predicted excess returns with the realized excess returns and sometimes predicts large negative excess returns, which is at odd with a positive reward expected  from taking risks.  The observed differences in the decomposition between estimates of  $a_{i,t}$ (orange shaded area) and of $b_{i,t}^{\top}\mathbb{E}[f_{t}\vert\mathcal{F}_{t-1}]$ (blue shaded area) come from the selected regressors in the first pass. Since the aLASSO penalization ends up with time-varying models presenting arbitrage, we observe larger values for estimated $\hat a_{i,t}$, especially during the recession periods (gray areas) determined by the National Bureau of Economic Research (NBER). The aOGL method avoids putting covariates in estimated $\hat a_{i,t}$
that should not be there because of the no-arbitrage restrictions. Besides, the estimated path for $a_{i,t}$ is close to zero with the aOGL method  as it should be if we believe that the factors are most of the time fully tradable.
\begin{table}[!ht]
\footnotesize
\centering
\begin{tabular}{@{}lcccccc@{}}
\toprule
  & \multicolumn{3}{c}{Carhart four-factor} & \multicolumn{3}{c}{Fama-French five-factor} \\
\cmidrule(lr){2-4}  \cmidrule(l){5-7}
  Methods & RMSPE & Av$(\vert \text{PE}\vert)$ & Std$(\vert\text{PE}\vert)$ & RMSPE & Av$(\vert \text{PE}\vert)$ & Std$(\vert\text{PE}\vert)$\\
 \cmidrule(r){1-1} \cmidrule(lr){2-2} \cmidrule(lr){3-3} \cmidrule(lr){4-4}  \cmidrule(lr){5-5}  \cmidrule(lr){6-6}  \cmidrule(l){7-7} 
 aOGL &  $1.46 \cdot 10^{-2}$ & $ 1.12 \cdot 10^{-2}$ & $ 0.93 \cdot 10^{-2}$  & $ 1.49 \cdot 10^{-2}$ & $ 1.16 \cdot 10^{-2}$ & $ 0.93 \cdot 10^{-2}$ \\
 aLASSO & $1.62 \cdot 10^{-2}$ & $ 1.27 \cdot 10^{-2}$ & $ 1.01 \cdot 10^{-2}$  & $ 2.14 \cdot 10^{-2}$ & $ 1.67 \cdot 10^{-2}$ & $ 1.35\cdot 10^{-2}$ \\
 TI &  $1.79 \cdot 10^{-2}$ & $ 1.36 \cdot 10^{-2}$ & $ 1.18 \cdot 10^{-2}$  & $ 1.37 \cdot 10^{-2}$ & $ 1.02 \cdot 10^{-2}$ & $ 0.91 \cdot 10^{-2}$ \\
 $\lambda_t \& \nu$ &  $1.45 \cdot 10^{-2}$ & $ 1.08 \cdot 10^{-2}$ & $ 0.97 \cdot 10^{-2}$ &  $1.46 \cdot 10^{-2}$ & $ 1.08 \cdot 10^{-2}$ & $ 0.98 \cdot 10^{-2}$ \\

\bottomrule
\end{tabular}
 \caption{Root Mean Squared Prediction Error (RMSPE), Mean Absolute Prediction Error (Av$(\vert \text{PE}\vert)$) and Standard Deviation  of the Absolute Prediction Error (Std$(\vert\text{PE}\vert)$) of an equally-weighted portfolio with the Carhart four-factor and Fama-French five-factor models  for the aOGL, aLASSO,  time-invariant (TI) and $\lambda_t \& \nu$ methods. The sample of US equity excess returns begins in July 1963 and ends in December 2019.}
\label{tab_rmspe_equal}
\end{table}
%

\subsection{Out-of-sample prediction performance}

\label{sec_perf_out}

In this section, we compare the out-of-sample prediction performance for the same methods used in the previous section. Here, we compute PE but for data that never enter into model estimation. We follow a similar approach to \cite{gu2020empirical}. We split the sample into two subsamples, one for training and one for testing. We estimate the models  from July 1963 to December 2009 and compute PE  from January 2010 to December 2019 (recent period). 
We repeat the same analysis for a training period from July 1963 to December 1999 and a testing period from January 2000 to December 2009 (older period). 
We closely follow the same setting as in the previous section, the only difference being that we separate the subsample used for estimation from the one used for prediction performance assessment.
\begin{table}[!ht]
\footnotesize
\centering
\begin{tabular}{@{}lcccccc@{}}
\toprule
& \multicolumn{6}{c}{Carhart four-factor} \\
\cmidrule(lr){2-7} 
 & \multicolumn{3}{c}{Jan.\ 2000 to Dec.\ 2009} & \multicolumn{3}{c}{Jan.\ 2010 to Dec.\ 2019} \\
\cmidrule(lr){2-4}  \cmidrule(l){5-7}
  Methods & RMSPE & Av$(\vert \text{PE}\vert)$ & Std$(\vert\text{PE}\vert)$ & RMSPE & Av$(\vert \text{PE}\vert)$ & Std$(\vert\text{PE}\vert)$\\
 \cmidrule(r){1-1} \cmidrule(lr){2-2} \cmidrule(lr){3-3} \cmidrule(lr){4-4}  \cmidrule(lr){5-5}  \cmidrule(lr){6-6}  \cmidrule(l){7-7} 
 aOGL &  $1.58 \cdot 10^{-2}$ & $ 1.23 \cdot 10^{-2}$ & $ 1.00 \cdot 10^{-2}$ &  $1.34 \cdot 10^{-2}$ & $ 1.06 \cdot 10^{-2}$ & $ 0.83 \cdot 10^{-2}$   \\
 aLASSO &  $2.43 \cdot 10^{-2}$ & $ 2.03 \cdot 10^{-2}$ & $ 1.37 \cdot 10^{-2}$ & $7.44 \cdot 10^{-2}$ & $ 6.17 \cdot 10^{-2}$ & $ 4.18  \cdot 10^{-2}$   \\
 TI &  $1.70 \cdot 10^{-2}$ & $ 1.32 \cdot 10^{-2}$ & $ 1.08 \cdot 10^{-2}$ &  $1.70 \cdot 10^{-2}$ & $ 1.32 \cdot 10^{-2}$ & $ 1.08 \cdot 10^{-2}$ \\
  $\lambda_t \& \nu$ &  $1.57 \cdot 10^{-2}$ & $ 1.24 \cdot 10^{-2}$ & $ 0.96 \cdot 10^{-2}$ &  $1.79 \cdot 10^{-2}$ & $ 1.31 \cdot 10^{-2}$ & $ 1.22 \cdot 10^{-2}$ \\

 \bottomrule
\end{tabular}
 \caption{Out-of-sample Root Mean Squared Prediction Error (RMSPE), Mean Absolute Prediction Error (Av$(\vert \text{PE}\vert)$) and Standard Deviation  of the Absolute Prediction Error (Std$(\vert\text{PE}\vert)$) of an equally-weighted portfolio with the Fama-French five-factor model for the aOGL, aLASSO, time-invariant (TI) and $\lambda_t \& \nu$ methods. The testing periods are Jan.\ 2000 to Dec.\ 2009 and Jan.\ 2010 to Dec.\ 2019. Their associated training periods precede them and start in July 1963. }
\label{tab_rmspe_equal_out_c4f}
\end{table}
\begin{table}[!ht]
\footnotesize
\centering
\begin{tabular}{@{}lcccccc@{}}
\toprule
& \multicolumn{6}{c}{Fama-French five-factor} \\
\cmidrule(lr){2-7}
 & \multicolumn{3}{c}{Jan.\ 2000 to Dec.\ 2009} & \multicolumn{3}{c}{Jan.\ 2010 to Dec.\ 2019} \\
\cmidrule(lr){2-4}  \cmidrule(l){5-7}
  Methods & RMSPE & Av$(\vert \text{PE}\vert)$ & Std$(\vert\text{PE}\vert)$ & RMSPE & Av$(\vert \text{PE}\vert)$ & Std$(\vert\text{PE}\vert)$\\
 \cmidrule(r){1-1} \cmidrule(lr){2-2} \cmidrule(lr){3-3} \cmidrule(lr){4-4}  \cmidrule(lr){5-5}  \cmidrule(lr){6-6}  \cmidrule(l){7-7} 
 aOGL &  $1.86 \cdot 10^{-2}$ & $ 1.38 \cdot 10^{-2}$ & $ 1.24 \cdot 10^{-2}$ &  $1.26 \cdot 10^{-2}$ & $ 0.99 \cdot 10^{-2}$ & $ 0.77 \cdot 10^{-2}$   \\
 aLASSO &  $9.05 \cdot 10^{-2}$ & $ 5.11 \cdot 10^{-2}$ & $ 7.49 \cdot 10^{-2}$ & $6.63 \cdot 10^{-2}$ & $ 5.99 \cdot 10^{-2}$ & $ 2.86  \cdot 10^{-2}$   \\
 TI &  $1.70 \cdot 10^{-2}$ & $ 1.32 \cdot 10^{-2}$ & $ 1.07 \cdot 10^{-2}$ &  $1.69 \cdot 10^{-2}$ & $ 1.32 \cdot 10^{-2}$ & $ 1.07 \cdot 10^{-2}$ \\
 $\lambda_t \& \nu$  &  $1.56 \cdot 10^{-2}$ & $ 1.24 \cdot 10^{-2}$ & $ 0.96 \cdot 10^{-2}$ &  $1.79 \cdot 10^{-2}$ & $ 1.32 \cdot 10^{-2}$ & $ 1.21 \cdot 10^{-2}$ \\
\bottomrule
\end{tabular}
 \caption{Out-of-sample Root Mean Squared Prediction Error (RMSPE), Mean Absolute Prediction Error (Av$(\vert \text{PE}\vert)$) and Standard Deviation  of the Absolute Prediction Error (Std$(\vert\text{PE}\vert)$) of an equally-weighted portfolio with the Fama-French five-factor model for the aOGL, aLASSO, time-invariant (TI) and $\lambda_t \& \nu$  methods. The testing periods are Jan.\ 2000 to Dec.\ 2009 and Jan.\ 2010 to Dec.\ 2019. Their associated training periods precede them and start in July 1963. }
\label{tab_rmspe_equal_out_ff5}
\end{table}
\begin{table}[!ht]
\small
\centering
\begin{tabular}{@{}ccccccccc@{}}
\toprule
& \multicolumn{8}{c}{Carhart four-factor} \\
\cmidrule(lr){2-9}
 & \multicolumn{4}{c}{Jan.\ 2000 to Dec.\ 2009} & \multicolumn{4}{c}{Jan.\ 2010 to Dec.\ 2019} \\
\cmidrule(lr){2-5}  \cmidrule(l){6-9}
  Year & aOGL & aLASSO  & TI &  $\lambda_t \& \nu$ & aOGL & aLASSO  & TI &  $\lambda_t \& \nu$\\
 \cmidrule(r){1-1} \cmidrule(lr){2-2} \cmidrule(lr){3-3} \cmidrule(lr){4-4}  \cmidrule(lr){5-5}  \cmidrule(lr){6-6}  \cmidrule(l){7-7} \cmidrule(l){8-8} 
 \cmidrule(l){9-9} 
  1 & 0.72 & 0.46 & 0.87 & 0.89 & 0.72 & 0.35 & 0.87 & 0.89   \\
  2 &  0.50 & 0.41 & 0.84 & 0.83 & 0.71 & 0.28 & 0.85 & 0.92   \\
  3 & 0.20 & 0.32 & 0.66 & 0.70 & 0.74 & 0.40 & 0.67 & 0.53   \\
 4  &  0.37 & 0.41 & 0.60 & 0.65 & 0.63 & 0.37 & 0.61 & 0.66   \\
  5 & 0.41 & 0.42 & 0.61 & 0.64 & 0.63 & 0.32 & 0.62 & 0.68   \\
  6 & 0.40 & 0.40 & 0.23 & 0.40 & 0.57 & 0.27 & 0.23 & 0.12   \\
  7 & 0.38 & 0.40 & 0.18 & 0.35 & 0.60 & 0.08 & 0.18 & 0.05   \\
 8 & 0.23 & 0.19 & 0.24 & 0.38 & 0.62 & -0.81 & 0.24 & 0.13  \\
 9 & -0.19 & -1.85 & 0.18 & 0.35 & 0.61 & -0.86 & 0.18 & 0.07   \\
\bottomrule
\end{tabular}
 \caption{Out-of-sample $R^2$ of an equally-weighted portfolio with the Carhart four-factor model for the aOGL, aLASSO, time-invariant (TI) and $\lambda_t \& \nu$ methods. The out-of-sample $R^2$ are computed for each year of the testing periods from Jan.\ 2000 to Dec.\ 2009 and from Jan.\ 2010 to Dec.\ 2019. Their associated training periods precede them and start in July 1963. }
\label{tab_r2_out_c4f}
\end{table}
\begin{table}[!ht]
\small
\centering
\begin{tabular}{@{}ccccccccc@{}}
\toprule
& \multicolumn{8}{c}{Fama-French four-factor} \\
\cmidrule(lr){2-9}
 & \multicolumn{4}{c}{Jan.\ 2000 to Dec.\ 2009} & \multicolumn{4}{c}{Jan.\ 2010 to Dec.\ 2019} \\
\cmidrule(lr){2-5}  \cmidrule(l){6-9}
  Year & aOGL & aLASSO  & TI &  $\lambda_t \& \nu$ & aOGL & aLASSO  & TI &  $\lambda_t \& \nu$\\
 \cmidrule(r){1-1} \cmidrule(lr){2-2} \cmidrule(lr){3-3} \cmidrule(lr){4-4}  \cmidrule(lr){5-5}  \cmidrule(lr){6-6}  \cmidrule(l){7-7} \cmidrule(l){8-8} 
 \cmidrule(l){9-9} 
  1 &0.69 &  0.32 & 0.86 & 0.89 & 0.72 & 0.25 & 0.87 & 0.87   \\
  2 & 0.50 & 0.31 & 0.84 &  0.83 & 0.74  & 0.20 & 0.84 & 0.91   \\
  3 & 0.16 & 0.17 & 0.66 & 0.70 & 0.79 & 0.07 & 0.66 & 0.50   \\
 4  & 0.35 & 0.19 & 0.60 & 0.65 & 0.67 & -0.06 & 0.60 & 0.65   \\
  5 & 0.39 & 0.22 & 0.61 & 0.65 & 0.65 & -0.20 & 0.61 & 0.68   \\
  6 & 0.38 & 0.23 & 0.23 & 0.40 & 0.60 & -0.29 & 0.23 & 0.11  \\
  7 & 0.35 & 0.23 & 0.18 & 0.35 & 0.64 & -0.57 & 0.18 & 0.05 \\
 8  & 0.30 & 0.38 & 0.24 & 0.38 & 0.60 & -0.19 & 0.24 & 0.13   \\
 9  & 0.09 & 0.30 & 0.18 & 0.34 & 0.58 & -1.17 & 0.18 & 0.07   \\
\bottomrule
\end{tabular}
 \caption{Out-of-sample $R^2$ of an equally-weighted portfolio with the Fama-French five-factor model for the aOGL, aLASSO, time-invariant (TI) and $\lambda_t \& \nu$ methods. The out-of-sample $R^2$ arew computed for each year of the testing periods from Jan.\ 2000 to Dec.\ 2009 and from Jan.\ 2010 to Dec.\ 2019.  Their associated training periods precede them and start in July 1963. }
\label{tab_r2_out_ff5}
\end{table}

\begin{figure}
\centering
 \includegraphics[width=1\linewidth]{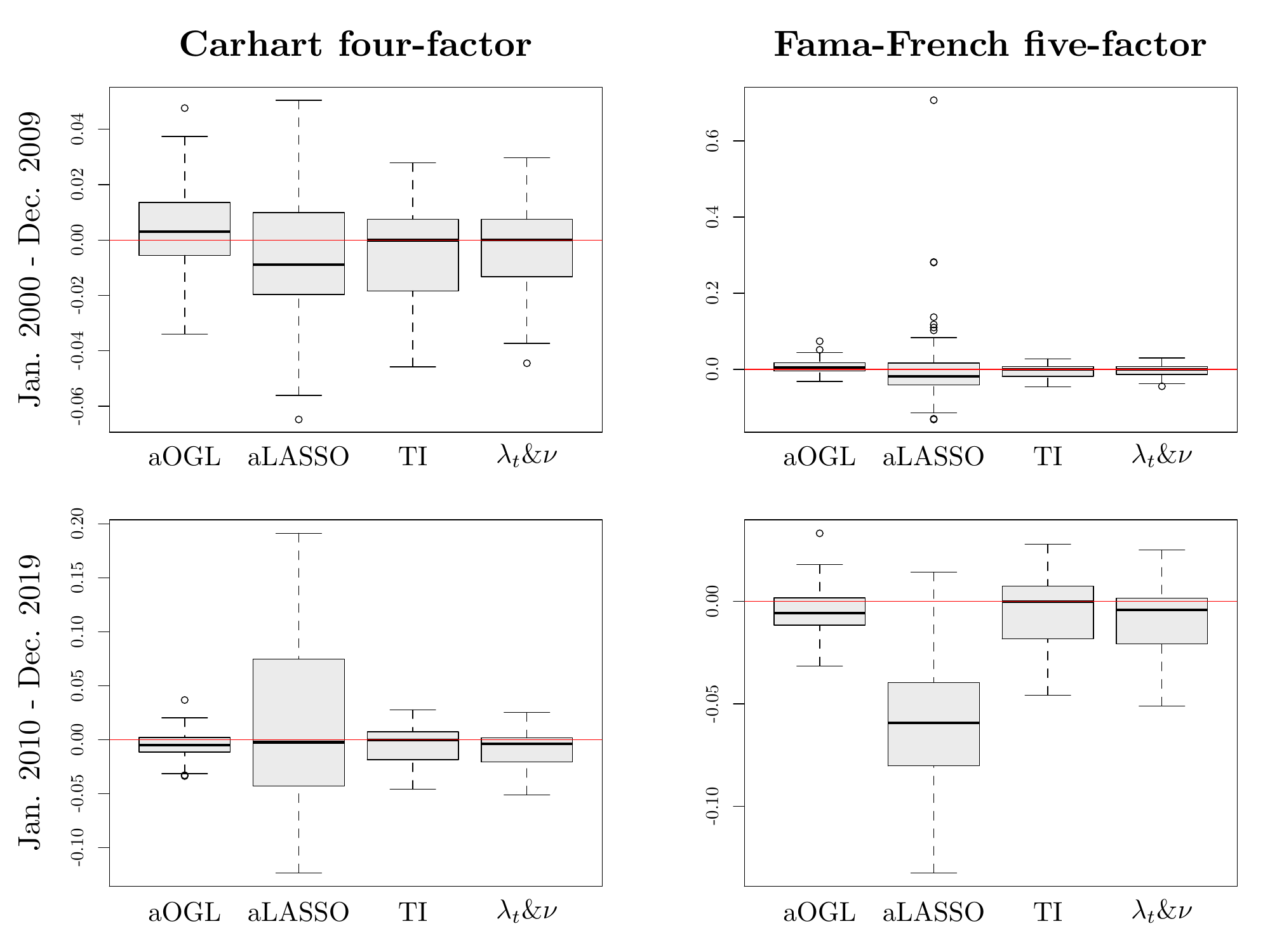}
   
 \caption{Empirical distribution of out-of-sample Prediction Error (PE) of an equally-weighted portfolio. We compare   the aOGL, aLASSO,  time-invariant (TI) and $\lambda_t \& \nu$ methods for the Carhart four-factor and Fama-French five-factor models. The upper panels are for the testing period 2000-2009. The lower panels are for the testing period 2010-2019.  Their associated training periods precede them and start in July 1963.}{}
 \label{fig_boxplot_out}
\end{figure}

We see that the aOGL method performs better than the aLASSO method in all cases as shown in Tables~\ref{tab_rmspe_equal_out_c4f} to \ref{tab_rmspe_equal_out_ff5}. Furthermore, the aOGL method often performs better than a time-invariant method  as exhibited by the RMSPE and the lower Std$(\vert\text{PE}\vert)$. Such an advantage over time-invariant alternatives is less clear  for the out-of-sample $R^2$ computed each year on the whole testing periods in Tables \ref{tab_r2_out_c4f} and \ref{tab_r2_out_ff5}. On the contrary, the aOGL method keeps a strong  advantage over the aLASSO method, especially for the years closer to the training periods.  
For both testing periods, the box-plots in Figure~\ref{fig_boxplot_out} show that out-of-sample PE related to the portfolio excess returns for the aOGL method are located closer to zero, more symmetrically distributed, and narrower.  As observed in the in-sample analysis, the aOGL method seems to perform better in terms of out-of-sample predictability as shown by the distributional behavior of the PE. We believe that the good out-of-sample performance for the portfolio comes from the diversification of the prediction errors among the single assets. We observe a similar phenomenon in forecast combinations \citep{Timmermann_2006}.

\section{Conclusions}
\label{sec_conclusion}

Our empirical results show that taking explicitly into account the no-arbitrage restrictions coming from the Arbitrage Pricing Theory do help in predictive modeling of large cross-sectional equity data sets with penalisation methods. We view this approach as an example of a structural approach to big data where incorporating finance theory improves on the prediction performance of the estimated quantities. It resonates with structural approaches in panel econometrics guided by economic theory \citep{bonhomme2017keeping}. In asset management and risk management, a better predictive performance of excess returns should help to better gauge time-variation in the risk-reward  trade-off.
In asset selection, it should help to improve performance of time-varying portfolio allocation when we use predicted excess returns as inputs. From our simulation and empirical results, we expect our procedure to perform  well in out-of-sample prediction  for portfolio building.

\section{Acknowledgements}

We would like to thank the Co-Editor, Associate Editor, and two referees for constructive criticism and numerous
suggestions which have led to substantial improvements over the previous version. We are grateful to I.\ Chaieb,
T.\ Chordia, A.-P.\ Fortin, P.\ Gagliardini, R.\ Garcia, M.\ Karemera, H.\ Langlois,   A.\ Patton, S.\ Pruitt, J.\ Thimme,  S.\ van den Hoff, D.\ Xiu, P.\ Zaffaroni for their helpful comments, as well as seminar participants at Queen Mary, University of Nottingham, Louvain Finance Seminar, IAE Lille, Universit\'e d'Orl\'eans, LUISS, Auburn University Statistics and Data Science Seminar, CREST-ENSAE Financial Econometrics Seminar,  and participants at the $8^{th}$ days of Econometrics for Finance, 2021 SoFiE Machine Learning conference, Vienna Workshop "Econometrics of Option Markets", 2021 SFI Research days, 2021 North America Summer Meeting of the Econometric Society, $13^{th}$ Annual SoFiE Conference, 2021 IAAE Conference, 2021 EcoSta, $7^{th}$ IYFS Conference, 2021 China International Conference in Finance (CICF), 2021 EEA-ESEM, $20^\text{th}$ Workshop in Econometrics for Finance, 2021 CFE-CM Statistics, $19^\text{th}$ Paris December Finance Meeting, ML approaches Finance and Management workshop, 38$^{th}$ AFFI conference, and QFFE conference 2022. G.\ Bakalli and O.\ Scaillet were supported by the SNSF Grant $\#100018-182582$. S.\ Guerrier was supported by the SNSF Professorships Grant \#176843 and by the Innosuisse-Boomerang Grant \#37308.1 IP-ENG.

\newpage
\clearpage

\newpage{}

\bibliographystyle{apalike}
\bibliography{ThesisBiB}

\newpage{}

\appendix
\section{Regularity conditions}
\label{append_regularity}

This Appendix lists and comments the regularity conditions needed to derive the asymptotic properties of the estimation procedure (see also Appendix A in \citetalias{gagliardini2016time}). Beforehand, recall the following vector $x_{i,t} =(\vech[X_t]^{\top}, \tilde{Z}_{t-1}^{\top} \otimes Z_{i,t-1}^{\top}, f_t^{\top} \otimes \tilde{Z}_{t-1}^{\top}, f_t^{\top} \otimes Z_{i,t-1}^{\top})^{\top}$ of dimension $d$.
\begin{Bassumption}\phantom{-}
 There exists a constant
$M$ such that  a) $\sup_i \Vert x_{i,t}\Vert \leq M$,
$P$-a.s.. Moreover, 
 b)  $\sup_i\Vert A_i\Vert<\infty$, $\sup_i\Vert B_i\Vert<\infty$, $\sup_i \Vert C_i \Vert<\infty$.
\label{Assum_Qxx}
\end{Bassumption}
\begin{Bassumption}\phantom{-}
\noindent $\inf_i\mathbbm{E}[I_{i,t}\vert \gamma_i]>0$ .
\label{assum_bound_observability}
\end{Bassumption}
\begin{Bassumption}\phantom{-}
\noindent $\inf_i \eig_{\min} ( \mathbbm{E}[x_{i,t}x_{i,t}^{\top}\vert \gamma_i])>0$, where $\eig_{\min}$ denotes the minimum eigenvalue of $\mathbbm{E}[x_{i,t}x_{i,t}^{\top}\vert \gamma_i]$.
\label{assum_bound_multicolinearity}
\end{Bassumption}
\begin{Bassumption}\phantom{-} 
The trimming constants satisfy
$\chi_{1,T}=\mathcal{O}\left((\log T)^{\kappa_{1}}\right)$, $\chi_{2,T}=\mathcal{O}\left((\log T)^{\kappa_{2}}\right)$,
with $\kappa_{1},\kappa_{2}>0$.
\label{Assum_CHI}
\end{Bassumption}

\begin{Bassumption}\phantom{-}
For $n$ sufficiently large, there exist sub-Gaussian random variables $Y_{i,l} \sim \subG(\sigma^2), \sigma^2 < \infty$ such that $\mathbb{E}[\max_{i,l} |\sqrt{T_i} (\hat{\beta}_{i,l} - \beta_{i,l})|] \leq \mathbb{E} [\max_{i,l} |Y_{i,l}|]$, for $i = 1, \ldots, n$, and $l = 1, \ldots, d$.
\label{assum_order_max}
\end{Bassumption}

\begin{Bassumption}\phantom{-}
We have that  $ \mathbbm{E}[u_t \vert u_{\underline{t-1}}, \mathcal{F}_t] = 0$ and there exists a constant $M>0$, such that $\Vert \mathbbm{E}[u_t u_t^{\top}\vert Z_{t-1}] \Vert \leq M$, for all $t$, where $u_t = f_t - \mathbbm{E}[f_t\vert\mathcal{F}_{t-1}]$.
\label{assum_bound_zt}
\end{Bassumption}

Assumption~\ref{Assum_Qxx} eases the proofs and requires uniform upper bounds on the regressor values, intercept, and model coefficients.  Assumption~\ref{assum_bound_observability} implies that the fraction of the time period in which an asset return is observed is bounded away from zero asymptotically  uniformly across assets, while Assumption~\ref{assum_bound_multicolinearity} bounds away from zero the minimum eigenvalue of the population squared moment to exclude asymptotic multicolinearity problems uniformly across assets. Assumption~\ref{Assum_CHI} gives an upper bound on the divergence rate of the trimming constants such that  logarithmic divergence rate allows to control the aOGL estimation error in the second-pass regression. Assumption~\ref{assum_order_max} is a technical requirement on $\mathbb{E}[\max_{i,l} |\sqrt{T_i} (\hat{\beta}_{i,l} - \beta_{i,l})|]$. From Lemma~\ref{lemma_asynorm_beta}, we have that $\hat{\beta}_{i,l}$ are asymptotically normally distributed and we might think that Assumption~\ref{assum_order_max} is directly satisfied due the properties of sub-Gaussian random variables. However, it is not the case due to our double asymptotics with $n, T \to \infty$. To illustrate the necessity of this requirement, we can consider the following example: let $Z_i \sim \mathcal{U}(0,1)$ and $\delta_{i,T} = i$  $\mathbf{1}_{i \geq T}$ for $i = 1, \ldots, n$. For all $i$, we have $X_{i,T} = Z_i + \delta_{i,T} \Longrightarrow \mathcal{U}(0,1)$ as $T \to \infty$. Suppose that $n > T$, then we have $\lim_{T \to \infty} \mathbb{E} [\max_{i=1,\ldots, n} |Z_{i}|] \leq 1$ while $\lim_{T \to \infty} \mathbb{E} [\max_{i=1,\ldots, n} |X_{i,T}|]$ diverges. We can replace Assumption~\ref{assum_order_max} by other requirements, for example, by considering sub-Gaussian error terms in the first-pass regression. Finally, Assumption~\ref{assum_bound_zt} allows for a martingale difference sequence and bounds the conditional variance-covariance matrix for the linear innovation  $u_t$ associated with the factor process. This assumption helps to prove consistency of the aLASSO estimator $\hat{F}_k$ using the same arguments as in Lemma~\ref{lemma_asynorm_beta}.

\vspace{.5cm}

\section{Proof of Lemma~\ref{lemma_asynorm_beta}}
\label{proof_lemma_asynorm_beta}

We follow the proof strategy of \cite{percival2012theoretical} (see \cite{nardi2008asymptotic} for related arguments for the Group-LASSO). Let $\beta^\star_{i} = \beta_i + \frac{u_i}{\sqrt{T_i}}$ and $\{v^\star_{i,g}\}$ and  $\{v_{i,g}\}$ be decomposition of $\beta_i$ minimizing $\|\beta^\star_i\|_{2,1,\groupset}$ and $\|\beta_i\|_{2,1,\groupset}$, respectively. Multiplying (\ref{eq_aogl}) by $\frac{T_i}{2}$, we have that
\begin{equation*}
    \mathcal{Q}^\star(u_i) = \frac{1}{2} \sum_t \left(I_{i,t} R_{i,t} - \left(\beta_i+\frac{u_i}{\sqrt{T_i}}\right)^\top I_{i,t} x_{i,t}\right)^2 + \delta  T_i \sum_g  \delta_g  \left\|v_{i,g} + \frac{1}{\sqrt{T_i}} v^{u_i}_{i,g} \right\|,
\end{equation*}
where $v^{u_i}_{i,g} = \sqrt{T_i}(v^\star_{i,g} - v_{i,g})$ is a decomposition of $u_i = \sqrt{T_i}(\beta^\star_{i} - \beta_i)$. We define 
$$\hat{u}_i = \argmin_{u_i\in\real^d} \mathcal{Q}^\star(u_i),$$
then we have $\hat{\beta}_{i} = \beta_i + \frac{\hat{u}_i}{\sqrt{T_i}}$. We write $D^\star(u_i) = \mathcal{Q}^\star(u_i) - \mathcal{Q}^\star(0)$ and thus we obtain
\begin{equation*}
    \begin{aligned}
        D^\star(u_i) &= \frac{1}{2}u_i^\top  \hat{Q}_{x,i} u_i - \frac{1}{\sqrt{T_i}} u_i^\top \sum_t I_{i,t}x_{i,t}\varepsilon_{i,t} \\
        &+ \sqrt{T_i} \delta \sum_g \delta_g \sqrt{T_i} \left(\left\|v_{i,g} + \frac{1}{\sqrt{T_i}} v^{u_i}_{i,g}\right\| - \left\|v_{i,g}\right\|\right) \\ 
        &= \mathcal{I}_{1} + \sum_g \mathcal{I}_{2,g}.
    \end{aligned}
\end{equation*}
From \cite{percival2012theoretical}, we know that, for $g \in G_{H_i}, \mathcal{I}_{2,g}$ vanishes to zero since $\delta_g$ based on an initial $\sqrt{T_i}$-consistent estimator goes to  ${\|v_{i,g}\|^{-\check{\gamma}}}$, from Assumption~\ref{assum:uniqueness}, the uniqueness of the decomposition of $v_{i,g}$ and $\sqrt{T_i}\delta = o(1)$. Moreover, for $g \in G_{H_i^c}, \mathcal{I}_{2,g}$ diverges  and, for $g \in G_{H_{0,i}}, \mathcal{I}_{2,g} $ diverges since $T_i ^{\nicefrac{\check{\gamma}}{2}}\|v_{i,g}^{\text{init}}\|^{\check{\gamma}} = \mathcal{O}_p(1)$ and $T_i^{\nicefrac{(1+\check{\gamma})}{2}}\delta$ diverges, where $v_{i,g}^{\text{init}}$ is the initial data dependent estimator of the latent decomposition of $\beta_i$. Moreover, under Assumption~\ref{assum:x_converge} and \ref{assum:mds}, using the CLT for martingale difference sequences and Slutsky's theorem, we have that 
\begin{equation*}
    \mathcal{I}_{1} \Longrightarrow \frac{1}{2} u_i^\top Q_{x,i} u_i - u_i^\top W_i, 
\end{equation*}
where $W_i \sim \mathcal{N}(0,\sigma^2_i Q_{x,i})$. It follows that 
$$ D^\star(u_i) \Longrightarrow D(u_i),$$ 
with 
\begin{equation*}
    D(u_i) = \left\{
		  \begin{array}{l l}
		    \frac{1}{2} u_{i}^\top Q_{x,i} u_{i} - u_{i}^\top W_i, \quad  \quad  \text{ if }  v^{u_i}_{i,g} \neq 0, \, \text{for}\,  g \in G_{H_i}, \\
			\infty, \quad \quad \quad \quad \, \, \, \quad \quad \quad \quad \quad \quad \quad  \quad  \text{ else}.
		  \end{array} \right.
\end{equation*}
Minimizing  $D(u_i)$ and using the argmax theorem from \cite{vandervaart1998weak} conclude the proof as in \cite{percival2012theoretical}. 
%
%
\begin{flushright}
$\square$
\end{flushright}

\section{Proof of Proposition  \ref{theo_consitency_nu}}
\label{proof_theo_consitency_nu}

From Lemma~\ref{lemma_asynorm_beta}, we have the convergence in distribution to a Gaussian random variable for all $i=1, \ldots, n$:
\begin{equation*}
    \sqrt{T_i}\left(\hat{\beta}_{i} - \beta_{i}\right) \Longrightarrow V_i.
\end{equation*}
%
%
%
%
%
%
Next, we consider the expectation of $\sup_{i} \boldsymbol{1}_{i}^{\chi} \big\| \hat{{\beta}_i} - {\beta}_i \big\|$. Fix $n$ to a sufficiently large value, then we have
\begin{equation*}
\begin{aligned}
        \mathbb{E}\left[\sup_{1 \leq i \leq n} \boldsymbol{1}_{i}^{\chi} \big\| \hat{{\beta}_i} - {\beta}_i \big\|\right] &= \mathbb{E}\left[\max_{1 \leq i \leq n} \boldsymbol{1}_{i}^{\chi} \sqrt{\sum_{l = 1}^d  \big( \hat{\beta}_{i,l} - \beta_{i,l} \big)^2}\right]\\
        & \leq \,\sqrt{d} \, \mathbb{E}\left[\max_{\substack{1 \leq i \leq n \\ 1 \leq l \leq d}} \boldsymbol{1}_{i}^{\chi} \big| \hat{\beta}_{i,l} - \beta_{i,l} \big|\right] \\
        &= \,  \frac{\sqrt{d}} {\sqrt{T}} \mathbb{E}\left[\max_{\substack{1 \leq i \leq n \\ 1 \leq l \leq d}} \boldsymbol{1}_{i}^{\chi} \sqrt{\frac{T}{T_i}} \sqrt{T_i}   \big| \hat{\beta}_{i,l} - \beta_{i,l} \big|\right]\\
         & \leq \,   \frac{\sqrt{d\, \chi_{2,T}}} {\sqrt{T}} \mathbb{E}\left[\max_{\substack{1 \leq i \leq n \\ 1 \leq l \leq d}} \sqrt{T_i}   \big| \hat{\beta}_{i,l} - \beta_{i,l} \big|\right].
\end{aligned}
\end{equation*}    
From Assumption~\ref{assum_order_max}, we have that 
\begin{equation*}
\begin{aligned}     
        \frac{\sqrt{d\, \chi_{2,T}}} {\sqrt{T}} \mathbb{E}\left[\max_{\substack{1 \leq i \leq n \\ 1 \leq l \leq d}} \sqrt{T_i}   \big| \hat{\beta}_{i,l} - \beta_{i,l} \big|\right] 
        &\quad \leq \, \frac{\sqrt{d \, \chi_{2,T} }}{\sqrt{T}}\,\,\left(\mathbb{E}\left[\max_{\substack{1 \leq i \leq n \\ 1 \leq l \leq d}} | Y_{i,l}| \right]\right) \\
        &\quad \leq \frac{\sqrt{2d \, \chi_{2,T} \, \sigma^2 \log(2nd) }}{\sqrt{T}}\,\,.
\end{aligned}
\end{equation*}
Thus, by Assumption~\ref{assum_wrong_model} and \ref{Assum_CHI}, there exist a positive constant $C$ such that
\begin{equation*}
\begin{aligned}
        \mathbb{E}\left[\sup_{1 \leq i \leq n} \boldsymbol{1}_{i}^{\chi} \big\| \hat{{\beta}_i} - {\beta}_i \big\|\right] \leq C \sqrt{\frac{ \log(T)^{1 + \kappa_2}}{T}},
\end{aligned}
\end{equation*}
and we have
\begin{equation*}
\begin{aligned}
        \lim_{n \to \infty}\;  \sqrt{\frac{T}{ \log(T)^{1 + \kappa_2}}} \mathbb{E}\left[\sup_{1 \leq i \leq n} \boldsymbol{1}_{i}^{\chi} \big\| \hat{{\beta}_i} - {\beta}_i \big\|\right] \leq C .
\end{aligned}
\end{equation*}
For $n$ sufficiently large, we have by Markov's inequality that for any $\epsilon > 0$
\begin{equation*}
\begin{aligned}
    &\Pr \left(\sqrt{\frac{ T}{\log(T)^{1+ \kappa_2}}}\sup_{1 \leq i \leq n} \boldsymbol{1}_{i}^{\chi} \big\| \hat{{\beta}}_i - {\beta}_i \big\|
    \geq \epsilon \right)  \\
    & \quad \leq \sqrt{\frac{ T}{\epsilon^2\log(T)^{1+ \kappa_2}}} \, \mathbb{E}\left[\sup_{1 \leq i \leq n} \boldsymbol{1}_{i}^{\chi} \big\| \hat{{\beta}_i} - {\beta}_i \big\|\right] \leq \frac{C}{\epsilon}.
\end{aligned}
\end{equation*}
Thus, we have 
\begin{equation*}
    \sup_{1 \leq i \leq n} \boldsymbol{1}_{i}^{\chi} \big\| \hat{{\beta}}_i - {\beta}_i \big\| = \mathcal{O}_p\left(\sqrt{\frac{ \log(T)^{1 + \kappa_2}}{T}}\right),
\end{equation*}
implying that 
\begin{equation*}
    \sup_{1 \leq i \leq n} \boldsymbol{1}_{i}^{\chi} \big\| \hat{{\beta}}_i - {\beta}_i \big\| = o_p(1).
\end{equation*}
The consistency of $\hat{\nu}$ then follows from the following results ii) $\sup_i\Vert w_i \Vert = O(1)$, iii) $\nicefrac{1}{n} \sum_i \Vert \hat{w}_i - w_i \Vert = o_p(1)$ and iv) $\hat{Q}_{\beta_3} - {Q}_{\beta_3} = o_p(1)$ of Lemma 3 of \citetalias{gagliardini2016time} under Assumptions~\ref{assum:x_converge} and \ref{Assum_Qxx} to \ref{Assum_CHI}. As in the proof of Proposition 3 of \citetalias{gagliardini2016time}, they ensure
$
 \Vert \hat{\nu} - {\nu} \Vert = o_p\left(1\right),
$
which concludes the proof.
\begin{flushright}
$\square$
\end{flushright}

\newpage

\section{No-arbitrage \textit{ex-ante} grouping structure}
\label{append_group}

This section is dedicated to describe how to construct the grouping structure needed to create the vector $\tilde{x}_{i,t}$ of duplicated regressors from the original $x_{i,t}$. We use the duplicated regressors  to implement the numerical optimisation of the aOGL method. From the set of Restrictions~\ref{res:ti} to \ref{res:zi}, it appears that, for any element in $x_{1,i,t}$ related to a specific element of $\tilde{Z}_{t-1,l}$ and $Z_{i,t-1,m}$, there exist multiple corresponding regressors in $x_{2,i,t}$ related to the same instrument $l$ and characteristic $m$. To implement a shrinkage estimator satisfying Restrictions~\ref{res:ti} to \ref{res:zi}, we  define the following  sets of indices.  The first group related to Restriction~\ref{res:ti}   always includes all covariates corresponding to the time-invariant contribution. Hence, we define $\tilde{x}^{(1)}_{i,t} = (x_{i,t,j})_{j \in \iota_{g_1}} \in \real^{n_1}$, where $n_1 = K+1$, and $\iota_{g_1}$ is a set of indices such that,
\begin{equation}
    \iota_{g_1} = \left\{1,d_1 + 1, \ldots, d_1 + k\tilde{p} + 1, \ldots, d_1+(K-1)\tilde{p} + 1\right\} \in \mathbb{N}_+^{K+1},
    \label{eq_g0}
\end{equation}
for $k= 1, \ldots, K-1$ and with $\mathbb{N}_+ = \natural \setminus \{0\}$.  The next set of groups are related to Restriction~\ref{res:no_group}, and we define $\tilde{x}^{(2)}_{i,t} = (x_{i,t,j})_{j \in \iota_{g_2}} \in \real^{n_2}$, where $n_2 = \tilde{p}(\tilde{p} - 1)/2 $, and
 the set $\iota_{g_2}$ corresponds to the indices related to the non-diagonal elements of $\vech(X_t)$ in $x_{i,t}$. To characterize it, let us first define the set of indices related to the diagonal elements in $\vech(X_t)$ (i.e., the squared elements $Z_{t-1,l}^2$) and the index set related to all elements in $\vech\left(X_t\right)$ as follows
\begin{equation*}
    \begin{aligned}
        \mathcal{D} & = \left\{x \in \natural_+ \left|\right. x = 1+(k-1)(\tilde{p}+1)-\frac{(k-1)k}{2}, k \in \{1,...,\tilde{p}\} \right\}, \\
        \mathcal{A} & =\left\{x \in \natural_+ |  x \leq \frac{(\tilde{p}+1)\tilde{p}}{2}\right\},
    \end{aligned}
\end{equation*}
such that the indices in $\mathcal{A} \backslash \mathcal{D}$ generate the set of indices: 
\begin{equation*}
    \iota_{g_2} = \{\iota_{g_{2,1}}, \ldots, \iota_{g_{2,n_{2}}} \}  \in \natural^{n_2}_+,
    \label{eq_g1}
\end{equation*}

Let us describe the group structure needed within a regular Group-LASSO by replicating our covariates to solve the original aOGL problem and ensuring that Restrictions~\ref{res:zt} and \ref{res:zi} are met. First, the scalar $u_l$, for $l = 1, \ldots, p$, denotes the $l$-th element of the set $ \mathcal{D}\setminus\{1\}$, i.e., the index set of diagonal elements excluding the first entry equal to 1, which  belongs already to $\iota_{g_1}$. Second, we duplicate  $K$ times each $u_l$ such that $u_{l,k}, k = 1, \ldots, K$, is the $k$-th duplicated element of $u_l$. Then, we can characterize   the set $\iota_{g_3}$ of indices  related to a scaled factor and its corresponding squared common instruments in the intercept as
\begin{equation}
    \iota_{g_3} = \{\iota_{g_{3,1}}, \ldots, \iota_{g_{3,Kp}} \}  \in \natural^{Kp}_+,
    \label{eq_g2}
\end{equation}
such that each set $\iota_{g_3,j} = \{u_{l,k},  d_1 +k + (l-1)\tilde{p} + 1\} \in \natural^{2}_+$, $k = 1, \ldots, K$, can generate a single group containing two covariates and $\tilde{x}^{(3)}_{i,t} = (x_{i,t,j})_{j \in \iota_{g_3}} \in \real^{n_3}$, where $n_3 = 2Kp$. Finally, the last set $\iota_{g_4}$ of indices  collects the indices related to Restrictions~\ref{res:zt} and \ref{res:zi} for the stock-specific instruments $Z_{i,t-1}$ such that 
\begin{equation}
        \iota_{g_4} = \{\iota_{g_{4,1}}, \ldots, \iota_{g_{4,Kq}} \}  \in \natural^{Kq}_+,
    \label{eq_g3}
\end{equation}
where each element $\iota_{g_{4,j}} = \{r_{m,k}, d_1 + d_{21} + k + (m-1)q + 1\} \in \natural^{\tilde{p}+1}_+, m = 1, \ldots, q$, $k = 1, ...,K$, and $r_{m,k}$ is the $k$-th duplicated set of indices 
\begin{equation*}
    r_{m,k} = \{d_{11} + m, \ldots, d_{11} + sq + m, \ldots,   d_{11} + pq + m\} \in \natural^{\tilde{p}+1}_+,
\end{equation*}
for $s = 1, \ldots, \tilde{p}$, $k=1, ...,K$. We define the last set of covariates groups as  $\tilde{x}^{(4)}_{i,t} = (x_{i,t,j})_{j \in \iota_{g_4}} \in \real^{n_4}$, where $n_4 = Kq(\tilde{p}+1)$. 
%
%
%
Next, we define the column vector
\begin{equation*}
    \tilde{x}_{i,t} = \left(\tilde{x}_{i,t}^{(1)\top} ,\tilde{x}_{i,t}^{(2)\top},\tilde{x}_{i,t}^{(3)\top},\tilde{x}_{i,t}^{(4)\top}\right)^\top \in \real^{\tilde{d}},
\end{equation*}
where $\tilde{d} = \sum_{j = 1}^4 n_j = K(\tilde{p}(q+2) + q -1) + (\tilde{p}-1)\tilde{p}/2 + 1 $. Let $\tilde{g} \in \widetilde{\mathcal{G}}$ denote a possible set of indices of the duplicated covariates $\tilde{x}_{i,t}$, where
\begin{equation}
     \widetilde{\mathcal{G}}
= \Big\{ \iota_{g_1}, \iota_{g_{2,1}}, \ldots, \iota_{g_{2,n_2}}, \iota_{g_{3,1}}, \ldots, \iota_{g_{3,Kp}}, \iota_{g_{4,1}} \ldots, \iota_{g_{4,Kq}} \Big\}.
\label{eq_tildeG}
\end{equation}
 The sets $\groupset$ and $\widetilde{\groupset}$ are based on the original covariates $x_{i,t}$ for the former and the duplicated covariates $\tilde{x}_{i,t}$ for the latter, and we have that $J = |\groupset| = |\widetilde{\groupset}| = 1 + n_2 +Kp+Kq$.

\end{document}